\documentclass[prl,twocolumn,floatfix,showpacs ]{revtex4-1}
\usepackage{amsmath}
\usepackage{amssymb}
\usepackage{graphicx}
\usepackage{dcolumn}
\usepackage{bm}
\usepackage[colorlinks=true,linkcolor=blue,anchorcolor=blue, citecolor=blue,urlcolor=blue]{hyperref}
\usepackage[mathlines]{lineno}
\usepackage{ulem}
\begin{document}
\title{High-temperature Majorana corner modes in a $d+id'$ superconductor heterostructure: Application to twisted bilayer cuprate superconductors}
\author{Yu-Xuan Li}
\author{Cheng-Cheng Liu}
\email{ccliu@bit.edu.cn}
\affiliation{Centre for Quantum Physics, Key Laboratory of Advanced Optoelectronic Quantum Architecture and Measurement (MOE), School of Physics, Beijing Institute of Technology, Beijing 100081, China}
\begin{abstract}
{The realization of Majorana corner modes generally requires unconventional superconducting pairing or $s$-wave pairing. However, the bulk nodes in  unconventional superconductors and the low $T_c$ of $s$-wave superconductors are not conducive to the experimental observation of Majorana corner modes.
Here we show the emergence of a Majorana corner mode at each corner of a two-dimensional topological insulator in proximity to a $d+id'$ pairing superconductor, such as heavily doped graphene or especially a twisted bilayer of a cuprate superconductor, e.g., Bi$_2$Sr$_2$CaCu$_2$O$_{8+\delta}$, which has recently been proposed as a fully gapped chiral $d_{x^2-y^2}+id_{xy}$ superconductor with $T_c$ close to its native 90 K, and an in-plane magnetic field.  By numerical calculation and intuitive edge theory, we find that the interplay of the proximity-induced pairing and Zeeman field can introduce opposite Dirac masses on adjacent edges of the topological insulator, which creates one zero-energy Majorana mode at each corner. Our scheme offers a feasible route to achieve and explore Majorana corner modes in a high-temperature platform without bulk superconductor nodes.}
\end{abstract}

\maketitle

\textit{\textcolor{black}{Introduction.---}}As the cornerstone of topological quantum computing, Majorana zero modes (MZMs) have attracted a lot of attention,  but searching MZMs is still a remarkable challenge in  quantum matter physics~\cite{re1,re2,re3,re4,re5,re6,re7,re8,re9,re10,RMPTSC,fu-Kane,re20,nanowire_Dassarm,re23,re14,re17,re18,re19,re24,re25,re26}. A nontrivial topological structure is essential for the creation of MZMs in topological superconductors (TSCs), which is usually characterized by  conventional bulk-boundary correspondence~\cite{RMPTSC}. Recently,  conventional TSCs have been generalized to their higher-order counterparts~\cite{benalcazar_quantized_2017,benalcazar_electric_2017,song2017,re50,schindler_2018,re51,yandw,wangspm,zhuso,re49,re39,liu_majorana_2018,yxwangprb,re53,RXzhang-iron,zhumix,yanodd,zccho,re42,pyati,cxliu-htsc,Fulga_prb,Piet_prx,re68,re43,bjyangtsc,re35,re41,sbzhangtsc,khhubbrd,ab-htsc,saha-Floquet,Shenchiral,qin_topological_2022,xjLiuprb,xxwu2}. In contrast to  conventional TSCs whose hallmark topological excitations are on the boundaries with co-dimension equal to one,  higher-order topological superconductors (HOTSCs) have protected topological characteristics on the boundaries with codimension greater than one. For example, two-dimensional (2D) HOTSCs could yield  unique 0D  MZMs localized at the corners of the sample, resulting in  Majorana corner modes (MCMs). Recently, some schemes for realizing MCMs have been proposed, in which the key component is the utilization of various superconductors, such as unconventional $d$-wave, $s_{\pm}$, as well as $p$-wave superconductors, and conventional $s$-wave superconductors~\cite{yandw,wangspm,zhuso,re43}. However, the bulk nodes in  unconventional superconductors and  the low $T_c$ of $s$-wave superconductors hinder the experimental detection of zero-energy MCMs.

Ever since the experimental discovery of correlated insulators and unconventional superconductivity in twisted bilayer graphene (TBG)~\cite{tbg1,tbg2}, the  twist as a new degree of freedom has opened up the new field of twistronics. Motivated by the novel phenomena of TBG and the experimental realization of 2D monolayer Bi$_2$Sr$_2$CaCu$_2$O$_{8+\delta}$ (Bi2212) with $T_c\approx 90$ K~\cite{scmo}, the concept of twistronics has also been extended to cuprate high-$T_c$ superconductors~\cite{scbi1,scbi3,scbi2,scbi4,scbi5,scbi6}. Recently, a twisted bilayer of high-$T_{c}$ cuprate monolayers at twist angle approaching $45^\circ$ was predicted as a fully gapped chiral $d_{x^2-y^2}+id_{xy}$ ($d+id^{\prime}$ for short) superconductor with the time-reversal symmetry (TRS) broken spontaneously up to temperatures approaching its native $T_c\approx 90 $ K~\cite{scbi1}. In addition, heavily doped graphene~\cite{nandkishore2012chiral,black-schaffer_edge_2012,black-schaffer_topological_2015}, bilayer silicene~\cite{liu_d_2013}, and a $\pi/2$ Josephson junction~\cite{yang__2018-1} were proposed to implement $d+id^\prime$ pairing. A natural question arises: Is it possible to realize MZMs by using a fully gapped high-$T_{c}$ $d+id^{\prime}$ superconductor so as to eliminate the above-mentioned disadvantages?

\begin{figure}
\includegraphics[scale=0.4]{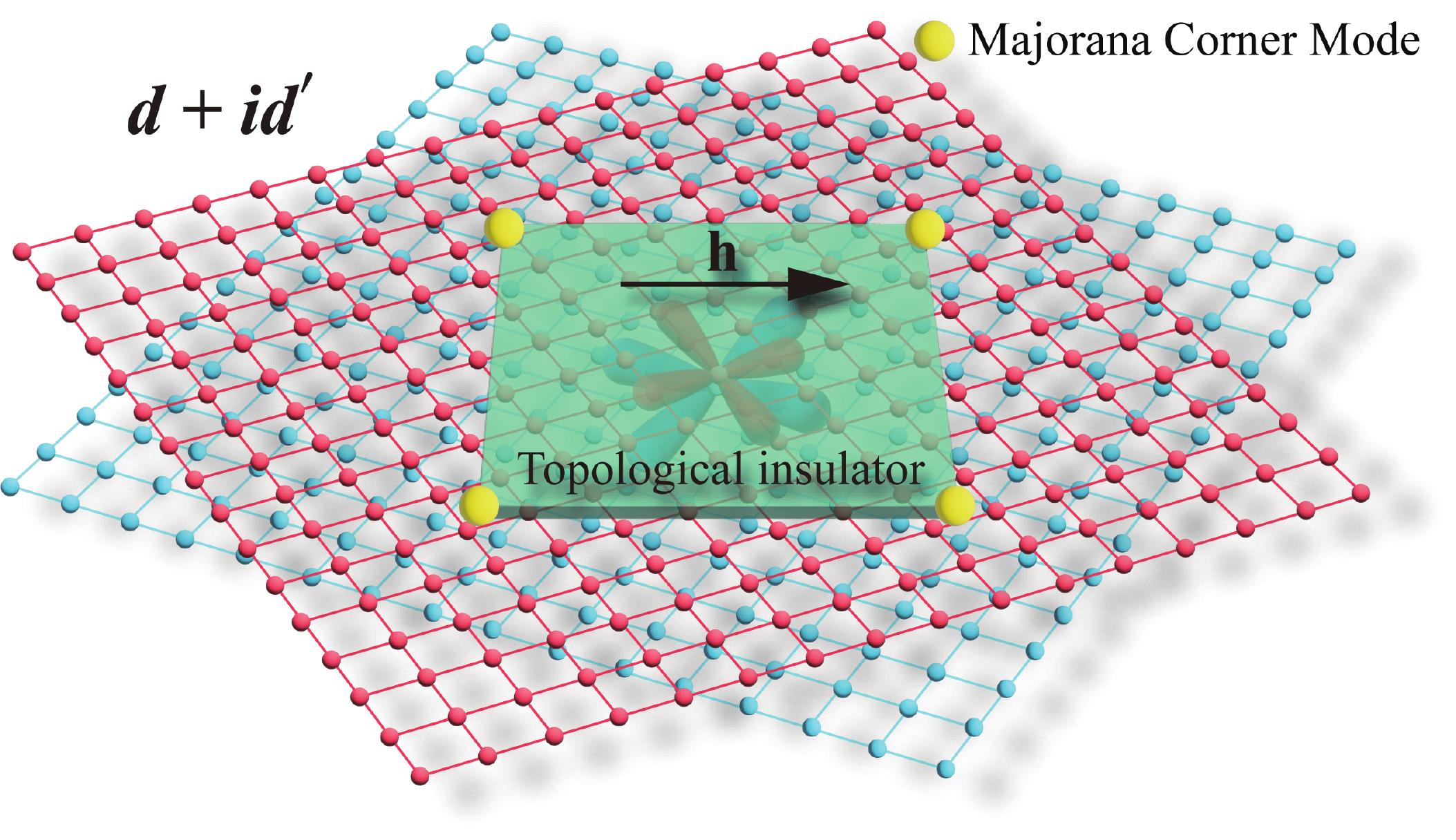}
\caption{Schematic diagram of the proposed setup. A heterostructure composed of a 2D topological insulator deposited on a high-$T_{c}$ fully gapped $d+id^{\prime}$ pairing superconductor such as a twisted bilayer of cuprate superconductor monolayers and subject to an in-plane Zeeman field. The sphere at each corner represents one zero-energy Majorana corner mode.}\label{f1}
\end{figure}

In this paper, we give the answer in the affirmative. We propose that MCMs can be achieved by growing a 2D topological insulator (TI) on twisted bilayer cuprate superconductors and imposing an in-plane Zeeman field, as illustrated in Fig.~\ref{f1}. The helical edge states of 2D TIs are protected by $U(1)$ gauge symmetry and TRS. The $d+id^{\prime}$ superconductor pairing induces a uniform Dirac mass for all the helical edge states, while an in-plane Zeeman field has contrasting effects along the different edges due to  spin-momentum locking in the helical edge states.  Beyond a critical Zeeman field, the resultant Dirac mass changes  sign at the corners, producing a zero-energy MCM at each corner as a mass-kink excitation.

\textit{\textcolor{black}{Physical system and minimal model.---}}We first introduce a heterostructure physical system consisting of a 2D TI (also known as a quantum spin Hall insulator) proximitized by a high-$T_{c}$ fully gapped $d+id^{\prime}$ twisted bilayer of cuprate superconductor (e.g., Bi$_2$Sr$_2$CaCu$_2$O$_{8+\delta}$) monolayers~\cite{supp} and an in-plane Zeeman field, as sketched in Fig.~\ref{f1}. We consider a minimal lattice model to describe the heterostructure, whose Bogoliubov$-$de Gennes (BdG) Hamiltonian is $\hat{H}=\sum_\mathbf{k}\Psi^\dagger_\mathbf{k}H^\text{BdG}(\mathbf{k})\Psi_\mathbf{k}$, with $\Psi^\dagger=(c^\dagger_{a\mathbf{k}\uparrow},c^\dagger_{b\mathbf{k}\uparrow},c^\dagger_{a\mathbf{k}\downarrow},c^\dagger_{b\mathbf{k}\downarrow},c_{a\mathbf{-k}\uparrow},c_{b\mathbf{-k}\uparrow},c_{a\mathbf{-k}\downarrow},c_{b\mathbf{-k}\downarrow})$ and
\begin{equation}
H^\text{BdG}(\mathbf{k})=\left(
\begin{array}{cc}
H(\mathbf{k})&\Delta(\mathbf{k})\\
-\Delta^*(\mathbf{-k})&-H^*(\mathbf{-k})
\end{array}
\right).\label{eq1}
\end{equation}
The normal state Hamiltonian is expressed as
\begin{equation}
\begin{aligned}
H(\mathbf{k})&=(m_0-t_x\cos k_x-t_y\cos k_y)\sigma_z\\
&+(\lambda_x\sin k_x s_y+\lambda_y\sin k_y s_x) \sigma_x+\mathbf{h}\cdot\mathbf{s}-\mu,
\end{aligned}\label{H_norm}
\end{equation}
 where $s_i$ and $\sigma_i$ are Pauli matrices denoting the electron spin $(\uparrow,\downarrow)$ and orbitals $(a,b)$, respectively. The first two terms make up the Hamiltonian for the 2D TI~\cite{bernevig_quantum_2006}, and the last two terms are the Zeeman term and the chemical potential. The proximity-induced $d+id'$ electron pairing can be expressed as
\begin{equation}
\Delta(\mathbf{k})=[\Delta_{1}(\cos k_{x}-\cos k_{y}) +i\Delta_{2}\sin k_{x}\sin k_{y}](-is_{y}).
\end{equation}
Throughout this work, $\lambda_{x,y}$, $t_{x,y}$, $\Delta_{1,2}$ are taken to be positive.
The 2D TI Hamiltonian is invariant under the space-inversion operation $\mathcal{I}=\sigma_z$ and TRS operation $\mathcal{T}=is_y\mathcal{K}$, where $\mathcal{K}$ is the complex-conjugation operator. When $m_0^2-(t_x+t_y)^2<0$ is satisfied, the Hamiltonian describes a 2D first-order TI in the band inverted region with the topological invariant $Z_2=1$ according to the parity criterion~\cite{fu_topological_2007}. The 2D first-order TI has gapless helical edge states protected by TRS and $U(1)$ gauge symmetry. In the presence of the Zeeman field and $d+id^{\prime}$ pairing, TRS and $U(1)$ gauge symmetry are both broken.

\textit{\textcolor{black}{Majorana corner modes.---}} On the one hand, due to the $U(1)$ gauge symmetry broken by  $d+id^{\prime}$ pairing, all the helical edge states of the TI are gapped out, confirmed by the direct calculation of the spectrum of the cylinder geometry~\cite{supp}, which introduces a homogenous Dirac mass for all the helical edge states. On the other hand, the spin-momentum locked helical edge states have different responses to the Zeeman field. For example, subject to Zeeman field $h_{x}$ ($h_{y}$) the helical edge states acquire a Dirac mass along the $k_{x}$ ($k_{y}$) direction while not along the $k_{y}$ ($k_{x}$). Without loss of generality, we first discuss the Zeeman field along the $x$ direction, with other in-plane directions and out-plane directions discussed later and in  the Supplemental Material~\cite{supp}.

\begin{figure}
\includegraphics[scale=0.4]{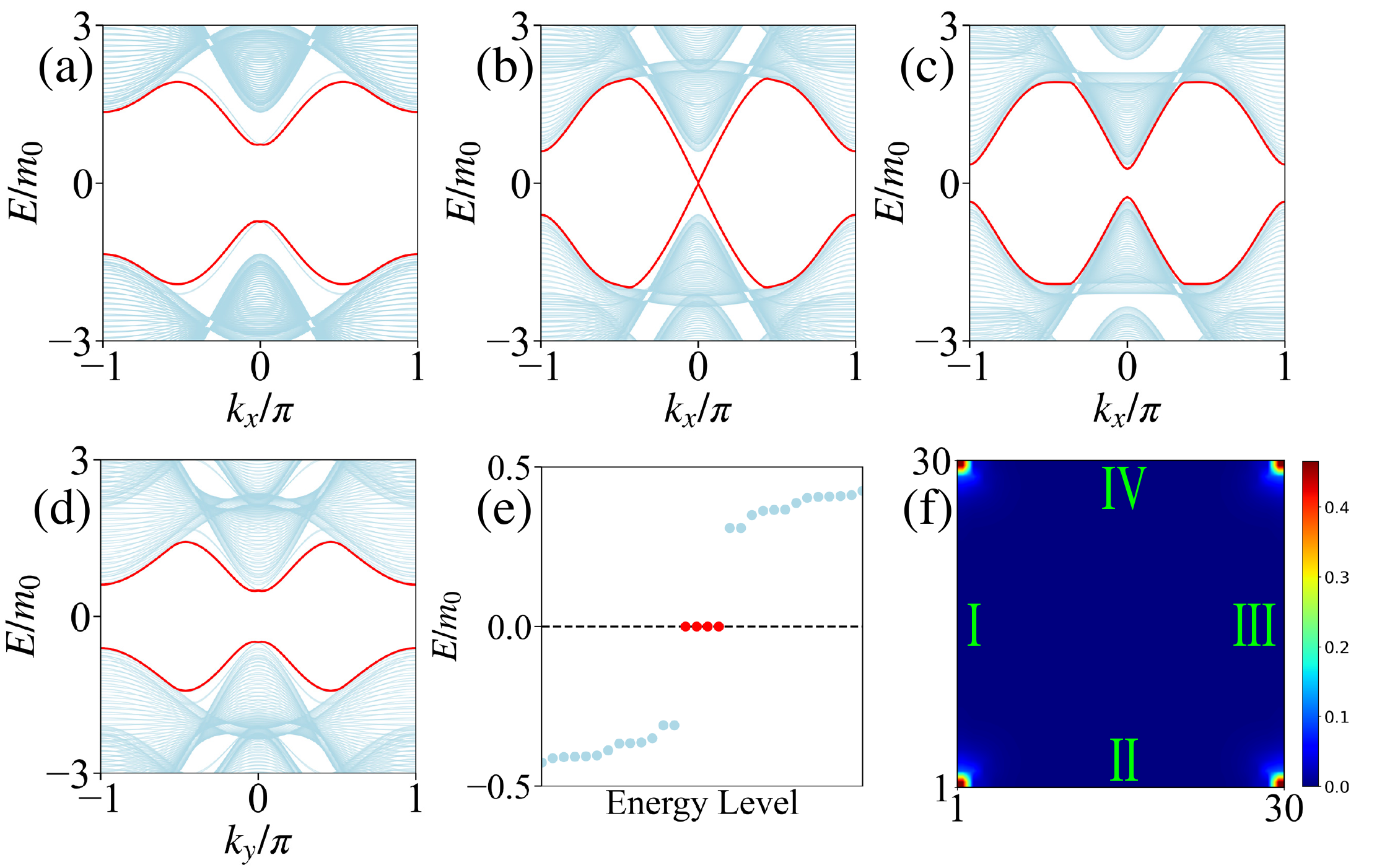}
\caption{Quasiparticle bands with edge spectra (red lines)  and bulk spectra (light blue lines) for an open boundary condition along the $y$ direction for (a) $h_{x}=0$, (b) $h_{x}=0.75$, and (c) $h_{x}=1$. The gap for the edge spectrum closes at the critical Zeeman field $h_{x}=0.75$. (d) Quasiparticle bands with the edge spectrum (red lines)  and bulk spectra (light blue lines) for open boundary conditions along the $x$ direction with the critical Zeeman field $h_{x}=0.75$. (e) Eigenvalues of the real-space TB Hamiltonian with $h_{x}=1$ for a 30$\times$30 square size sample. (f) The density plot displays the corner localized probability distribution of the four zero-energy MCMs in (e). I, II, III, and IV label the four edges. Common parameters are $m_{0}=1, t_{x}=t_{y}=\lambda_{x}=\lambda_{y}=2, \Delta_{1}=\Delta_{2}=0.5, \mu=0.1.$}\label{f2}
\end{figure}

 With the increase of the Zeeman field, the spectrum of the edge states will undergo a closing and reopening evolution along the $k_x$ direction, while along the $k_y$ direction the edge states are always gapped, as plotted in Figs.~\ref{f2}(a)$-$\ref{f2}(d), indicating the existence of one  edge-localized mass domain with the in-plane Zeeman field exceeding the critical value $h_{x}^c$ .  Note that throughout this transition, there is no gap closing in the bulk band. Such an edge-localized mass domain would give rise to MCMs and drive the system to a second-order TSC phase. In order to confirm such an intuitive scenario, we directly perform numerical calculations of the energy spectra of a square sample in the topological regime, as shown in Fig.~\ref{f2}(e). In spite of the fact that the 2D bulk states and the 1D edge states are all fully gapped, four MZMs emerge in the energy spectra of the nano-flake sample. The corresponding wave function distribution of the four MZMs in real space is exhibited in Fig.~\ref{f2}(f) with one MZM at each corner, namely MCMs.

\textit{\textcolor{black}{Edge theory.---}}To provide a better understanding of the emergence of MCMs, we derive the low-energy theory on each edge. To simplify the picture, we take $\mu=0$ and focus on the continuum model by expanding the lattice Hamiltonian in Eq.~(\ref{eq1}) to $O(k^2)$  around $\mathbf{k}=(0,0)$,
\begin{equation}
\begin{aligned}
	H_{\text{eff}}(\mathbf{k})&=(m+\frac{t_x}{2}k_x^2+\frac{t_y}{2}k_y^2)\sigma_z\tau_z+\lambda_xk_x\sigma_xs_y\tau_{z}\\
	&+\lambda_yk_y\sigma_xs_{x}-\frac{\Delta_1}{2}(k_x^2-k_y^2)s_y\tau_y+\Delta_2k_xk_ys_y\tau_x\\
	&+h_{x}s_{x}\tau_{z},
\end{aligned}
\end{equation}
where $m=m_0-t_x-t_y<0$ is satisfied to guarantee that the normal state has helical edge states in the topological nontrivial phase. We consider a semi-infinite geometry occupying the space $x\ge0$ for edge I [Fig.~\ref{f2}(f)]. We can replace $k_x$ by $-i\partial_x$ and divide the Hamiltonian as $H=H_{0}+H_p$ with $H_{0}(-i\partial_x,k_y)=(m-t_x\partial_x^2/2)\sigma_z\tau_z-i\lambda_x\sigma_xs_y\tau_{z}\partial_x$, and $H_p(-i\partial_x,k_y)=\lambda_yk_y\sigma_xs_{x}+\Delta_1/2s_y\tau_y\partial_x^2-i\Delta_2k_y\partial_xs_y\tau_x+h_{x}s_{x}\tau_{z}$, where all the $k_y^2$ terms are omitted. Solving the zero-energy solutions of $H_{0}\psi_\alpha(x)=0$ with the boundary condition $\psi_{\alpha}(0)=\psi_{\alpha}(+\infty)=0$, we find four zero-energy solutions, whose eigenstates are $\psi_\alpha(x)=\mathcal{N}_x\sin(\kappa_1x)e^{-\kappa_2x}e^{ik_yy}\xi_\alpha$,
where $\kappa_1=\sqrt{|(2m/t_x)|-(\lambda_x^2/t_x^2)}, \kappa_2=(\lambda_x/t_x)$,  and $\mathcal{N}_x$ is the normalization constant. The eigenvectors $\xi_\alpha$ satisfy $\sigma_ys_y\xi_\alpha=-\xi_\alpha$.
 Here we choose $\xi_{\alpha, \alpha=1-4}=|\tau_z\rangle\otimes|\sigma_ys_y=-1\rangle$~\cite{supp}. In the bases of the four eigenstates, the matrix elements of $H_{p}$ read~\cite{supp}
\begin{equation}
H_{\text{I}}=\lambda_{y}k_{y}{\eta}_{x}-M_{\text{I}}\tau_{y}\eta_{z},
\end{equation}
where ${\eta}$ are Pauli matrices in the two bases of $\chi_1=|\sigma_y=+1\rangle\otimes |s_y=-1\rangle, \chi_2=|\sigma_y=-1\rangle\otimes |s_y=+1\rangle$,
and the Dirac mass $M_\text{I}$ is equal to $-\Delta_{1}|m|/t_{x}$~\cite{supp}.

Similarly, the low-energy effective Hamiltonian for edge II, edge III, and edge IV can be obtained~\cite{supp}. The Dirac masses generated from the superconductor pairing for other edges are $M_\text{III}=M_\text{I}$, and $M_\text{II}=M_\text{IV}=\Delta_1|m|/t_y$. To facilitate the discussion, we can define "edge coordinate" $l$, which changes in a counterclockwise direction, so that the low-energy edge theory can be uniformly expressed as
\begin{equation}
H_{\text{edge},l}=i\lambda(l)\eta_x\partial_l-M(l)\tau_{y}\eta_{z}-h(l)\eta_z,\label{edgeeff}
\end{equation}
where $\lambda(l)=\{\lambda_y,\lambda_x,\lambda_y,\lambda_x\}$, $M(l)=\{-\Delta_1|m|/t_x, \Delta_1|m|/t_y, -\Delta_1|m|/t_x, \Delta_1|m|/t_y\}$ and $h(l)=\{0,h_{x},0,h_{x}\}$ for $l=$ \{I-IV\}, respectively. We find that edge I and edge III have one Dirac mass from the electron pairing, while edge II and  edge IV have two Dirac masses from the competing electron pairing and Zeeman field terms. The low-energy edge spectra are $E_\text{I,III}(k_y)=\pm\sqrt{(\lambda_{y}k_y)^2+M_\text{I,III}^2}$ and $E_\text{II,IV}(k_x)=\pm\sqrt{(\lambda_{x}k_x)^2+(M_\text{II,IV}\pm h_{x})^2}$.
 As $h_{x}$ increases, the gap along the $k_x$ boundary first closes at the critical Zeeman field $h^c_x\equiv M_\text{II}$ and reopens when $h_{x}>M_\text{II}$. The boundary topology phase transition takes place along the $k_x$ direction at $h^c_x$, while along the $k_y$ direction the edge gap is always fully gapped  which is consistent with the direct numerical calculations, as shown in Figs.~\ref{f2}(a)$-$\ref{f2}(d).

To explain the physics of the MCMs more clearly, we decouple the edge Hamiltonian Eq.~(\ref{edgeeff}) as $H_{\text{edge},l}=H_{\text{edge},l}^+\oplus H_{\text{edge},l}^-$ according to $\tau_{y}=\pm1$. When $h_{x}>h^c_x$, the mass terms of each edge  for  $H_{\text{edge},l}^-$ have the same signs; however, the mass terms of the four edges for  $H_{\text{edge},l}^+$ change with alternating signs. The effective edge Hamiltonian  for the $\tau_{y}=+1$ section reads
\begin{equation}
H^{+}_{\text{edge},l}=i\lambda(l)\eta_x\partial_l+\tilde{M}(l)\eta_z,
\end{equation}
with the mass $\tilde{M}(l=\text{I-IV})=\{\Delta_{1}|m|/t_{x},-\Delta_1|m|/t_y-h_{x},\Delta_{1}|m|/t_{x},-\Delta_1|m|/t_y-h_{x}\}$. One can see the signs of mass of adjacent edges are opposite,  i.e., $\tilde{M}(l)\tilde{M}(l\pm1)<0$, resulting in the emergence of one MCM at each corner.

\begin{figure}
\includegraphics[scale=0.35]{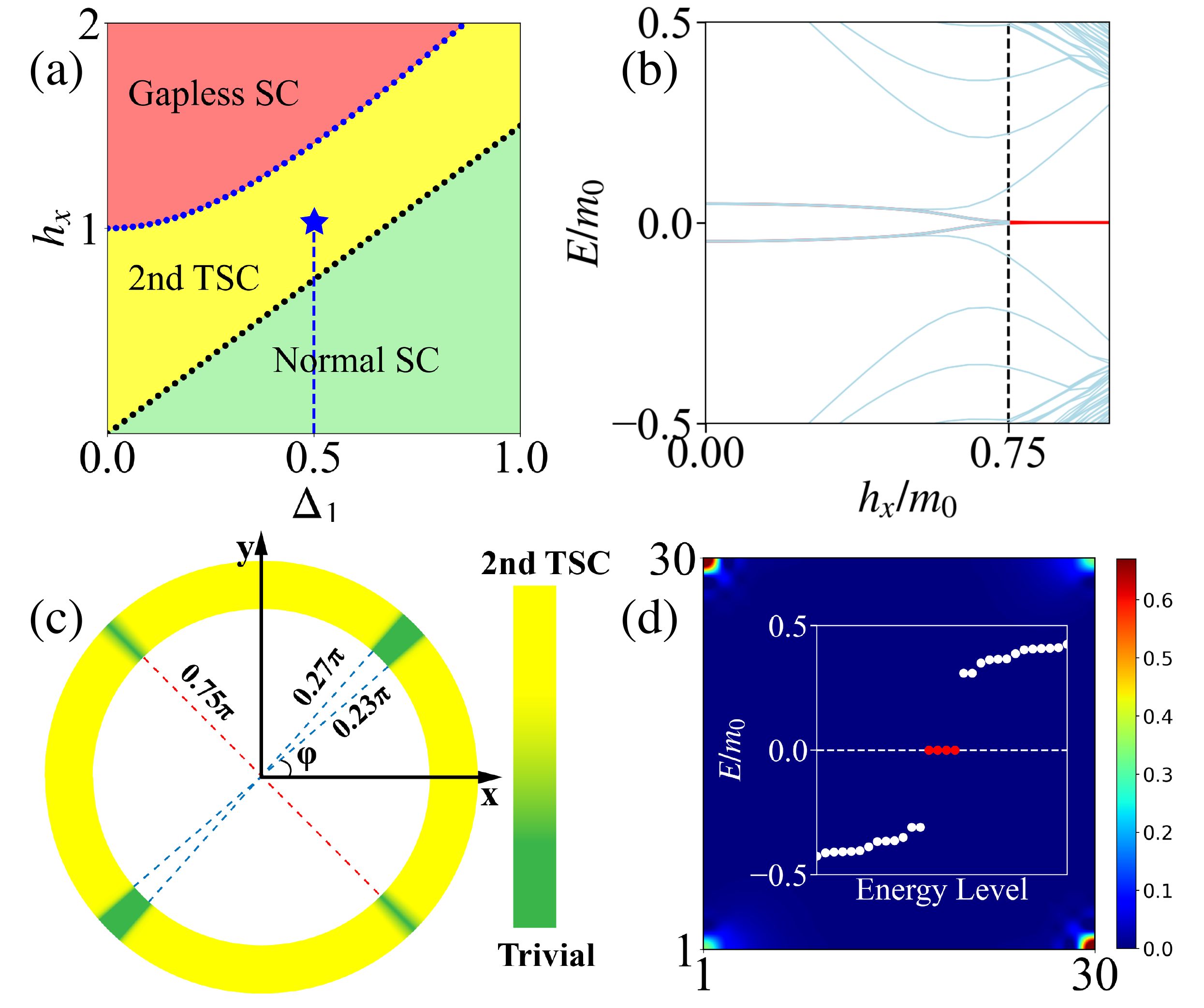}
\caption{(a) Phase diagram vs $h_{x}$ and $\Delta_{1}$. (b) Line-scan real-space spectra along the blue dashed line in (a). $\mu=0.1$. (c) Phase diagram as a function of the azimuth of the in-plane Zeeman field. Parameters are $h_\parallel=m_{0},h_{x}=h_{\parallel}\cos\varphi,h_{y}=h_{\parallel}\sin \varphi,\mu=0.$ (d) Energy spectrum of the real-space TB Hamiltonian at $\varphi=0.4\pi, \mu=0.1$ for a 30 $\times$ 30 square size sample.  The density plot exhibits the corner localized probability distribution of the four zero-energy MCMs. Common parameters are $m_0=1, \Delta_{1}=\Delta_{2}=0.5, t_x= t_y= \lambda_x=\lambda_y =2.$}\label{f3}
\end{figure}

\textit{\textcolor{black}{Phase diagram.---}} When the applied external Zeeman field is gradually increased, the gaps of both the bulk states and the edge states parallel to the Zeeman field decrease simultaneously, as shown in Figs.~\ref{f2}(a)$-$\ref{f2}(c). In this process, the gap of the edge states is closed first, and then the gap of the bulk states is closed again. We have known from the edge theory that when the Zeeman field $h_{x}=h^c_x\equiv M_\text{II}$, the gap of the edge states will close, and the system starts to enter into the second-order TSC phase. The bulk spectra have a simple expression $E(\mathbf{k})=\pm\sqrt{\lambda^2_x\sin k_x^2+(\sqrt{\epsilon^2(\mathbf{k})}\pm h_x)^2}$ with $\mu =0$, where $\epsilon^2(\mathbf{k})\equiv M^2(\mathbf{k})+\lambda^2_y\sin k_y^2+\Delta_1^2(\mathbf{k})+\Delta_2^2(\mathbf{k})$  with $M(\mathbf{k})\equiv m_{0}-t_{x}\cos k_{x}-t_{y}\cos k_{y}$. As we continue to increase the Zeeman field, the gap of the bulk state will close at $\mathbf{X} (\pi, 0)$ with $h_{x}=\sqrt{(m_0+t_x-t_y)^2+4\Delta^2_1}\equiv h_{x}^\mathbf{X}$. As a result, we analytically obtain the parameter interval of the phase diagram
\begin{equation}
		\left\{
		\begin{aligned}
			0<&h_{x}<M_\text{II},\quad\text{Normal SC}\\
M_\text{II}<&h_{x}<h_{x}^\mathbf{X},\quad\text{2nd TSC}\\
h_{x}>&h_{x}^\mathbf{X}\qquad\qquad\text{Gapless SC}.\\
		\end{aligned}
		\right.
\end{equation}
We present the phase diagram versus $h_{x}$ and $\Delta_{1}$ in Fig.~\ref{f3}(a), where the second-order TSC (2nd TSC) exists between the normal superconductor (normal SC) and gapless superconductor (gapless SC) phases and the phase boundary is illustrated in black and blue dotted lines. We further calculate the energy spectrum of a nano flake sample as shown in Fig.~\ref{f3}(b), with the pairing $\Delta_{1}$ fixed but varying the Zeeman field size as signaled by the blue dashed line in Fig.~\ref{f3}(a). The zero-energy MCMs emerge once the Zeeman field exceeds the critical value $h_{x}^c$, which is consistent with our edge theory.

So far, we have considered the $x$-direction Zeeman field. Now we turn to the general in-plane Zeeman field $\mathbf{h}=h_{\parallel}(\cos\varphi, \sin\varphi)$ with $\varphi$ as the azimuth, and obtain the effective edge Hamiltonian~\cite{supp}, which reads
\begin{equation}
H^{\rm eff}_{\text{edge},l}=i\lambda(l)\eta_x\partial_l-M(l)\tau_{y}\eta_{z}-h(l,\varphi)\eta_z,
\end{equation}
with $h(l=\text{I-IV},\varphi)=h_{\parallel}\{\sin\varphi, \cos\varphi, \sin\varphi, \cos\varphi\}$. Similarly, the Hamiltonian $H^{\rm eff}_{\text{edge},l}$ can be decoupled into two sub-blocks according to $\tau_{y}=\pm1$. Obviously, the Dirac masses of the edges depend on azimuth $\varphi$.  The emergence of one stable MCM at each corner requires that the signs of Dirac masses of the adjacent edges for one sub-block are opposite while for the other the signs are identical.  We calculate the evolution for the signs of the Dirac masses versus the azimuth for each sub-block in the Supplemental Material~\cite{supp}.  Accordingly, we find the requirement is met when the azimuth $\varphi$ is outside the interval $[0.23\pi,0.27\pi]\cup[1.23\pi,1.27\pi]$~\cite{supp}. In addition, the rotation of the in-plane Zeeman field may lead to the closure of the edge gap at the critical azimuth $\varphi^{c}=0.75\pi/1.75\pi$~\cite{supp}, where MCMs do not exist. Consequently, we obtain the phase diagram versus azimuth $\varphi$, as illustrated in Fig.~\ref{f3}(c), in which the large yellow areas mark a second-order TSC phase with the hallmark MCMs. We take $\varphi=0.4\pi$ as an example and calculate the energy spectrum and wave-function distribution for a nano flake sample as shown in Fig.~\ref{f3}(d), with one zero-energy MCM localized at each corner.

\textit{\textcolor{black}{Discussion and conclusion.---}} Our proposal is experimentally feasible because the necessary ingredients and the required technology are already available. Single-layer Bi$_2$Sr$_2$CaCu$_2$O$_{8+\delta}$ (Bi2212) superconductors have been grown~\cite{scmo}, and a twisted system of several layers of Bi2212 has also been fabricated~\cite{XueTwsited_d-wave2021,zhao_emergent_2021}, which encourage further endeavors to thin down the Bi2212 twisted systems to the monolayer limit.  Recently, graphene has been successfully overdoped beyond the van Hove singularity experimentally~\cite{rosenzweig_overdoping_2020}, which provides an unprecedented opportunity to access the $d+id'$ pairing. On the other hand, the quantum spin Hall effect has been experimentally observed in monolayer WTe$_{2}$ at 100K \cite{wu_observation_2018} and near room temperature in bismutene~\cite{reis_bismuthene_2017}. Superconductivity-induced meV-level pairing gaps at the boundary states of topological insulators have been detected in many topological insulator/superconductor heterostructure systems~\cite{lupke_proximity-induced_2020,shimamura_ultrathin_2018,wang_fully_2013,xu_artificial_2014,zhao_superconducting_2018}, such as a 0.7 meV pairing gap in WTe$_{2}$/NbSe$_{2}$~\cite{lupke_proximity-induced_2020}, a 7.5 meV pairing gap in bilayer-Bi/Bi2212~\cite{shimamura_ultrathin_2018}, a 15 meV pairing gap in Bi$_{2}$Se$_{3}$/Bi2212~\cite{wang_fully_2013}, etc. Among these diversiform topological insulator/superconductor heterostructures, we might as well choose a setup consisting of monolayer WTe$_{2}$ in proximity to the twisted bilayer Bi2212. Although the specific value of the pairing gap in monolayer WTe$_{2}$ induced by proximitized twisted bilayer Bi2212 has not been reported yet, estimates on the order of 1 meV should be reasonable. The large Land\'{e} $g$ factor (4.5$-$44) of WTe$_{2}$~\cite{aivazian_magnetic_2015,bi_spin_2018,wu_observation_2018}, which depends on the direction of magnetic field, enables an external magnetic field on the order of 1 T to induce a suitable Zeeman effect with the emergence of MCMs.

Scanning tunneling microscopy (STM) can be used to detect and resolve the spatial profile of the zero bias peaks induced by the MCM localized at the sample corner~\cite{re70}.  In a quantum point contact, the MCM can induce resonant Andreev reflection with a quantized zero-bias conductance peak of $2e^2/h$~\cite{law_majorana_2009,wimmer_quantum_2011}. One can build a superconductor-superconductor (S-S) junction where two corners are in contact and a finite phase difference is allowed between superconductors. Such an S-S junction with two corners in contact may host a coupled Majorana pair. The pair of Majoranas can mediate a fractional Josephson effect~\cite{re3} and crossed Andreev reflection~\cite{nilsson_splitting_2008}. Taken together, these four methods would provide compelling evidence of the existence of the MCMs.

In conclusion, we have demonstrated that a heterostructure composed of topological insulators and twisted bilayer cuprate superconductors can host MCMs when an in-plane Zeeman field is applied. Our proposed setup with fully gap pairing and high transition temperature has great advantages for the experimental observation of the zero-energy MCM signals. Our work may also stimulate further studies of MCMs in twisted systems.

It is our pleasure to thank Fan Yang and Yugui Yao for insightful discussions. The work is supported by National Key R\&D Program of China (Grant No. 2020YFA0308800) and the NSF of China (Grant No. 11922401).
%


\begin{thebibliography}{92}%
    \makeatletter
    \providecommand \@ifxundefined [1]{%
        \@ifx{#1\undefined}
    }%
    \providecommand \@ifnum [1]{%
        \ifnum #1\expandafter \@firstoftwo
        \else \expandafter \@secondoftwo
        \fi
    }%
    \providecommand \@ifx [1]{%
        \ifx #1\expandafter \@firstoftwo
        \else \expandafter \@secondoftwo
        \fi
    }%
    \providecommand \natexlab [1]{#1}%
    \providecommand \enquote  [1]{``#1''}%
    \providecommand \bibnamefont  [1]{#1}%
    \providecommand \bibfnamefont [1]{#1}%
    \providecommand \citenamefont [1]{#1}%
    \providecommand \href@noop [0]{\@secondoftwo}%
    \providecommand \href [0]{\begingroup \@sanitize@url \@href}%
    \providecommand \@href[1]{\@@startlink{#1}\@@href}%
    \providecommand \@@href[1]{\endgroup#1\@@endlink}%
    \providecommand \@sanitize@url [0]{\catcode `\\12\catcode `\$12\catcode
        `\&12\catcode `\#12\catcode `\^12\catcode `\_12\catcode `\%12\relax}%
    \providecommand \@@startlink[1]{}%
    \providecommand \@@endlink[0]{}%
    \providecommand \url  [0]{\begingroup\@sanitize@url \@url }%
    \providecommand \@url [1]{\endgroup\@href {#1}{\urlprefix }}%
    \providecommand \urlprefix  [0]{URL }%
    \providecommand \Eprint [0]{\href }%
    \providecommand \doibase [0]{http://dx.doi.org/}%
    \providecommand \selectlanguage [0]{\@gobble}%
    \providecommand \bibinfo  [0]{\@secondoftwo}%
    \providecommand \bibfield  [0]{\@secondoftwo}%
    \providecommand \translation [1]{[#1]}%
    \providecommand \BibitemOpen [0]{}%
    \providecommand \bibitemStop [0]{}%
    \providecommand \bibitemNoStop [0]{.\EOS\space}%
    \providecommand \EOS [0]{\spacefactor3000\relax}%
    \providecommand \BibitemShut  [1]{\csname bibitem#1\endcsname}%
    \let\auto@bib@innerbib\@empty
    \bibitem [{\citenamefont {Read}\ and\ \citenamefont {Green}(2000)}]{re1}%
    \BibitemOpen
    \bibfield  {author} {\bibinfo {author} {\bibfnamefont {N.}~\bibnamefont
            {Read}}\ and\ \bibinfo {author} {\bibfnamefont {D.}~\bibnamefont {Green}},\
    }\href {\doibase 10.1103/PhysRevB.61.10267} {\bibfield  {journal} {\bibinfo
            {journal} {Phys. Rev. B}\ }\textbf {\bibinfo {volume} {61}},\ \bibinfo
        {pages} {10267} (\bibinfo {year} {2000})}\BibitemShut {NoStop}%
    \bibitem [{\citenamefont {Volovik}(1999)}]{re2}%
    \BibitemOpen
    \bibfield  {author} {\bibinfo {author} {\bibfnamefont {G.~E.}\ \bibnamefont
            {Volovik}},\ }\href {\doibase 10.1134/1.568223} {\bibfield  {journal}
        {\bibinfo  {journal} {JETP Lett.}\ }\textbf {\bibinfo {volume} {70}},\
        \bibinfo {pages} {609} (\bibinfo {year} {1999})}\BibitemShut {NoStop}%
    \bibitem [{\citenamefont {Kitaev}(2001)}]{re3}%
    \BibitemOpen
    \bibfield  {author} {\bibinfo {author} {\bibfnamefont {A.~Y.}\ \bibnamefont
            {Kitaev}},\ }\href {\doibase 10.1070/1063-7869/44/10s/s29} {\bibfield
        {journal} {\bibinfo  {journal} {Phys. Usp.}\ }\textbf {\bibinfo {volume}
            {44}},\ \bibinfo {pages} {131} (\bibinfo {year} {2001})}\BibitemShut
    {NoStop}%
    \bibitem [{\citenamefont {Beenakker}(2013)}]{re4}%
    \BibitemOpen
    \bibfield  {author} {\bibinfo {author} {\bibfnamefont {C.}~\bibnamefont
            {Beenakker}},\ }\href {\doibase 10.1146/annurev-conmatphys-030212-184337}
    {\bibfield  {journal} {\bibinfo  {journal} {Annu. Rev. Condens. Matter
                Phys.}\ }\textbf {\bibinfo {volume} {4}},\ \bibinfo {pages} {113} (\bibinfo
        {year} {2013})}\BibitemShut {NoStop}%
    \bibitem [{\citenamefont {Alicea}(2012)}]{re5}%
    \BibitemOpen
    \bibfield  {author} {\bibinfo {author} {\bibfnamefont {J.}~\bibnamefont
            {Alicea}},\ }\href {\doibase 10.1088/0034-4885/75/7/076501} {\bibfield
        {journal} {\bibinfo  {journal} {Rep. Prog. Phys.}\ }\textbf {\bibinfo
            {volume} {75}},\ \bibinfo {pages} {076501} (\bibinfo {year}
        {2012})}\BibitemShut {NoStop}%
    \bibitem [{\citenamefont {Stanescu}\ and\ \citenamefont {Tewari}(2013)}]{re6}%
    \BibitemOpen
    \bibfield  {author} {\bibinfo {author} {\bibfnamefont {T.~D.}\ \bibnamefont
            {Stanescu}}\ and\ \bibinfo {author} {\bibfnamefont {S.}~\bibnamefont
            {Tewari}},\ }\href {\doibase 10.1088/0953-8984/25/23/233201} {\bibfield
        {journal} {\bibinfo  {journal} {J. Phys. Condens. Matter}\ }\textbf {\bibinfo
            {volume} {25}},\ \bibinfo {pages} {233201} (\bibinfo {year}
        {2013})}\BibitemShut {NoStop}%
    \bibitem [{\citenamefont {Leijnse}\ and\ \citenamefont
        {Flensberg}(2012)}]{re7}%
    \BibitemOpen
    \bibfield  {author} {\bibinfo {author} {\bibfnamefont {M.}~\bibnamefont
            {Leijnse}}\ and\ \bibinfo {author} {\bibfnamefont {K.}~\bibnamefont
            {Flensberg}},\ }\href {\doibase 10.1088/0268-1242/27/12/124003} {\bibfield
        {journal} {\bibinfo  {journal} {Semicond. Sci. Technol.}\ }\textbf {\bibinfo
            {volume} {27}},\ \bibinfo {pages} {124003} (\bibinfo {year}
        {2012})}\BibitemShut {NoStop}%
    \bibitem [{\citenamefont {Elliott}\ and\ \citenamefont {Franz}(2015)}]{re8}%
    \BibitemOpen
    \bibfield  {author} {\bibinfo {author} {\bibfnamefont {S.~R.}\ \bibnamefont
            {Elliott}}\ and\ \bibinfo {author} {\bibfnamefont {M.}~\bibnamefont
            {Franz}},\ }\href {\doibase 10.1103/RevModPhys.87.137} {\bibfield  {journal}
        {\bibinfo  {journal} {Rev. Mod. Phys.}\ }\textbf {\bibinfo {volume} {87}},\
        \bibinfo {pages} {137} (\bibinfo {year} {2015})}\BibitemShut {NoStop}%
    \bibitem [{\citenamefont {Sarma}\ \emph {et~al.}(2015)\citenamefont {Sarma},
        \citenamefont {Freedman},\ and\ \citenamefont {Nayak}}]{re9}%
    \BibitemOpen
    \bibfield  {author} {\bibinfo {author} {\bibfnamefont {S.~D.}\ \bibnamefont
            {Sarma}}, \bibinfo {author} {\bibfnamefont {M.}~\bibnamefont {Freedman}}, \
        and\ \bibinfo {author} {\bibfnamefont {C.}~\bibnamefont {Nayak}},\ }\href
    {\doibase 10.1038/npjqi.2015.1} {\bibfield  {journal} {\bibinfo  {journal}
            {npj Quantum Inf.}\ }\textbf {\bibinfo {volume} {1}},\ \bibinfo {pages}
        {15001} (\bibinfo {year} {2015})}\BibitemShut {NoStop}%
    \bibitem [{\citenamefont {Sato}\ and\ \citenamefont {Fujimoto}(2016)}]{re10}%
    \BibitemOpen
    \bibfield  {author} {\bibinfo {author} {\bibfnamefont {M.}~\bibnamefont
            {Sato}}\ and\ \bibinfo {author} {\bibfnamefont {S.}~\bibnamefont
            {Fujimoto}},\ }\href {\doibase 10.7566/JPSJ.85.072001} {\bibfield  {journal}
        {\bibinfo  {journal} {J. Phys. Soc. Jpn.}\ }\textbf {\bibinfo {volume}
            {85}},\ \bibinfo {pages} {072001} (\bibinfo {year} {2016})}\BibitemShut
    {NoStop}%
    \bibitem [{\citenamefont {Qi}\ and\ \citenamefont {Zhang}(2011)}]{RMPTSC}%
    \BibitemOpen
    \bibfield  {author} {\bibinfo {author} {\bibfnamefont {X.-L.}\ \bibnamefont
            {Qi}}\ and\ \bibinfo {author} {\bibfnamefont {S.-C.}\ \bibnamefont {Zhang}},\
    }\href {\doibase 10.1103/RevModPhys.83.1057} {\bibfield  {journal} {\bibinfo
            {journal} {Rev. Mod. Phys.}\ }\textbf {\bibinfo {volume} {83}},\ \bibinfo
        {pages} {1057} (\bibinfo {year} {2011})}\BibitemShut {NoStop}%
    \bibitem [{\citenamefont {Fu}\ and\ \citenamefont {Kane}(2008)}]{fu-Kane}%
    \BibitemOpen
    \bibfield  {author} {\bibinfo {author} {\bibfnamefont {L.}~\bibnamefont
            {Fu}}\ and\ \bibinfo {author} {\bibfnamefont {C.~L.}\ \bibnamefont {Kane}},\
    }\href {\doibase 10.1103/PhysRevLett.100.096407} {\bibfield  {journal}
        {\bibinfo  {journal} {Phys. Rev. Lett.}\ }\textbf {\bibinfo {volume} {100}},\
        \bibinfo {pages} {096407} (\bibinfo {year} {2008})}\BibitemShut {NoStop}%
    \bibitem [{\citenamefont {Oreg}\ \emph {et~al.}(2010)\citenamefont {Oreg},
        \citenamefont {Refael},\ and\ \citenamefont {von Oppen}}]{re20}%
    \BibitemOpen
    \bibfield  {author} {\bibinfo {author} {\bibfnamefont {Y.}~\bibnamefont
            {Oreg}}, \bibinfo {author} {\bibfnamefont {G.}~\bibnamefont {Refael}}, \ and\
        \bibinfo {author} {\bibfnamefont {F.}~\bibnamefont {von Oppen}},\ }\href
    {\doibase 10.1103/PhysRevLett.105.177002} {\bibfield  {journal} {\bibinfo
            {journal} {Phys. Rev. Lett.}\ }\textbf {\bibinfo {volume} {105}},\ \bibinfo
        {pages} {177002} (\bibinfo {year} {2010})}\BibitemShut {NoStop}%
    \bibitem [{\citenamefont {Lutchyn}\ \emph {et~al.}(2010)\citenamefont
        {Lutchyn}, \citenamefont {Sau},\ and\ \citenamefont
        {Das~Sarma}}]{nanowire_Dassarm}%
    \BibitemOpen
    \bibfield  {author} {\bibinfo {author} {\bibfnamefont {R.~M.}\ \bibnamefont
            {Lutchyn}}, \bibinfo {author} {\bibfnamefont {J.~D.}\ \bibnamefont {Sau}}, \
        and\ \bibinfo {author} {\bibfnamefont {S.}~\bibnamefont {Das~Sarma}},\ }\href
    {\doibase 10.1103/PhysRevLett.105.077001} {\bibfield  {journal} {\bibinfo
            {journal} {Phys. Rev. Lett.}\ }\textbf {\bibinfo {volume} {105}},\ \bibinfo
        {pages} {077001} (\bibinfo {year} {2010})}\BibitemShut {NoStop}%
    \bibitem [{\citenamefont {Cook}\ and\ \citenamefont {Franz}(2011)}]{re23}%
    \BibitemOpen
    \bibfield  {author} {\bibinfo {author} {\bibfnamefont {A.}~\bibnamefont
            {Cook}}\ and\ \bibinfo {author} {\bibfnamefont {M.}~\bibnamefont {Franz}},\
    }\href {\doibase 10.1103/PhysRevB.84.201105} {\bibfield  {journal} {\bibinfo
            {journal} {Phys. Rev. B}\ }\textbf {\bibinfo {volume} {84}},\ \bibinfo
        {pages} {201105} (\bibinfo {year} {2011})}\BibitemShut {NoStop}%
    \bibitem [{\citenamefont {Zhang}\ \emph {et~al.}(2013)\citenamefont {Zhang},
        \citenamefont {Kane},\ and\ \citenamefont {Mele}}]{re14}%
    \BibitemOpen
    \bibfield  {author} {\bibinfo {author} {\bibfnamefont {F.}~\bibnamefont
            {Zhang}}, \bibinfo {author} {\bibfnamefont {C.~L.}\ \bibnamefont {Kane}}, \
        and\ \bibinfo {author} {\bibfnamefont {E.~J.}\ \bibnamefont {Mele}},\ }\href
    {\doibase 10.1103/PhysRevLett.111.056402} {\bibfield  {journal} {\bibinfo
            {journal} {Phys. Rev. Lett.}\ }\textbf {\bibinfo {volume} {111}},\ \bibinfo
        {pages} {056402} (\bibinfo {year} {2013})}\BibitemShut {NoStop}%
    \bibitem [{\citenamefont {Nadj-Perge}\ \emph {et~al.}(2014)\citenamefont
        {Nadj-Perge}, \citenamefont {Drozdov}, \citenamefont {Li}, \citenamefont
        {Chen}, \citenamefont {Jeon}, \citenamefont {Seo}, \citenamefont {MacDonald},
        \citenamefont {Bernevig},\ and\ \citenamefont {Yazdani}}]{re17}%
    \BibitemOpen
    \bibfield  {author} {\bibinfo {author} {\bibfnamefont {S.}~\bibnamefont
            {Nadj-Perge}}, \bibinfo {author} {\bibfnamefont {I.~K.}\ \bibnamefont
            {Drozdov}}, \bibinfo {author} {\bibfnamefont {J.}~\bibnamefont {Li}},
        \bibinfo {author} {\bibfnamefont {H.}~\bibnamefont {Chen}}, \bibinfo {author}
        {\bibfnamefont {S.}~\bibnamefont {Jeon}}, \bibinfo {author} {\bibfnamefont
            {J.}~\bibnamefont {Seo}}, \bibinfo {author} {\bibfnamefont {A.~H.}\
            \bibnamefont {MacDonald}}, \bibinfo {author} {\bibfnamefont {B.~A.}\
            \bibnamefont {Bernevig}}, \ and\ \bibinfo {author} {\bibfnamefont
            {A.}~\bibnamefont {Yazdani}},\ }\href {\doibase 10.1126/science.1259327}
    {\bibfield  {journal} {\bibinfo  {journal} {Science}\ }\textbf {\bibinfo
            {volume} {346}},\ \bibinfo {pages} {602} (\bibinfo {year}
        {2014})}\BibitemShut {NoStop}%
    \bibitem [{\citenamefont {Jeon}\ \emph {et~al.}(2017)\citenamefont {Jeon},
        \citenamefont {Xie}, \citenamefont {Li}, \citenamefont {Wang}, \citenamefont
        {Bernevig},\ and\ \citenamefont {Yazdani}}]{re18}%
    \BibitemOpen
    \bibfield  {author} {\bibinfo {author} {\bibfnamefont {S.}~\bibnamefont
            {Jeon}}, \bibinfo {author} {\bibfnamefont {Y.}~\bibnamefont {Xie}}, \bibinfo
        {author} {\bibfnamefont {J.}~\bibnamefont {Li}}, \bibinfo {author}
        {\bibfnamefont {Z.}~\bibnamefont {Wang}}, \bibinfo {author} {\bibfnamefont
            {B.~A.}\ \bibnamefont {Bernevig}}, \ and\ \bibinfo {author} {\bibfnamefont
            {A.}~\bibnamefont {Yazdani}},\ }\href {\doibase 10.1126/science.aan3670}
    {\bibfield  {journal} {\bibinfo  {journal} {Science}\ }\textbf {\bibinfo
            {volume} {358}},\ \bibinfo {pages} {772} (\bibinfo {year}
        {2017})}\BibitemShut {NoStop}%
    \bibitem [{\citenamefont {Kim}\ \emph {et~al.}(2018)\citenamefont {Kim},
        \citenamefont {Palacio-Morales}, \citenamefont {Posske}, \citenamefont
        {Rózsa}, \citenamefont {Palotás}, \citenamefont {Szunyogh}, \citenamefont
        {Thorwart},\ and\ \citenamefont {Wiesendanger}}]{re19}%
    \BibitemOpen
    \bibfield  {author} {\bibinfo {author} {\bibfnamefont {H.}~\bibnamefont
            {Kim}}, \bibinfo {author} {\bibfnamefont {A.}~\bibnamefont
            {Palacio-Morales}}, \bibinfo {author} {\bibfnamefont {T.}~\bibnamefont
            {Posske}}, \bibinfo {author} {\bibfnamefont {L.}~\bibnamefont {Rózsa}},
        \bibinfo {author} {\bibfnamefont {K.}~\bibnamefont {Palotás}}, \bibinfo
        {author} {\bibfnamefont {L.}~\bibnamefont {Szunyogh}}, \bibinfo {author}
        {\bibfnamefont {M.}~\bibnamefont {Thorwart}}, \ and\ \bibinfo {author}
        {\bibfnamefont {R.}~\bibnamefont {Wiesendanger}},\ }\href {\doibase
        10.1126/sciadv.aar5251} {\bibfield  {journal} {\bibinfo  {journal} {Sci.
                Adv.}\ }\textbf {\bibinfo {volume} {4}},\ \bibinfo {pages} {eaar5251}
        (\bibinfo {year} {2018})}\BibitemShut {NoStop}%
    \bibitem [{\citenamefont {Zhang}\ \emph {et~al.}(2018)\citenamefont {Zhang},
        \citenamefont {Yaji}, \citenamefont {Hashimoto}, \citenamefont {Ota},
        \citenamefont {Kondo}, \citenamefont {Okazaki}, \citenamefont {Wang},
        \citenamefont {Wen}, \citenamefont {Gu}, \citenamefont {Ding},\ and\
        \citenamefont {Shin}}]{re24}%
    \BibitemOpen
    \bibfield  {author} {\bibinfo {author} {\bibfnamefont {P.}~\bibnamefont
            {Zhang}}, \bibinfo {author} {\bibfnamefont {K.}~\bibnamefont {Yaji}},
        \bibinfo {author} {\bibfnamefont {T.}~\bibnamefont {Hashimoto}}, \bibinfo
        {author} {\bibfnamefont {Y.}~\bibnamefont {Ota}}, \bibinfo {author}
        {\bibfnamefont {T.}~\bibnamefont {Kondo}}, \bibinfo {author} {\bibfnamefont
            {K.}~\bibnamefont {Okazaki}}, \bibinfo {author} {\bibfnamefont
            {Z.}~\bibnamefont {Wang}}, \bibinfo {author} {\bibfnamefont {J.}~\bibnamefont
            {Wen}}, \bibinfo {author} {\bibfnamefont {G.~D.}\ \bibnamefont {Gu}},
        \bibinfo {author} {\bibfnamefont {H.}~\bibnamefont {Ding}}, \ and\ \bibinfo
        {author} {\bibfnamefont {S.}~\bibnamefont {Shin}},\ }\href {\doibase
        10.1126/science.aan4596} {\bibfield  {journal} {\bibinfo  {journal}
            {Science}\ }\textbf {\bibinfo {volume} {360}},\ \bibinfo {pages} {182}
        (\bibinfo {year} {2018})}\BibitemShut {NoStop}%
    \bibitem [{\citenamefont {Wang}\ \emph
        {et~al.}(2018{\natexlab{a}})\citenamefont {Wang}, \citenamefont {Kong},
        \citenamefont {Fan}, \citenamefont {Chen}, \citenamefont {Zhu}, \citenamefont
        {Liu}, \citenamefont {Cao}, \citenamefont {Sun}, \citenamefont {Du},
        \citenamefont {Schneeloch}, \citenamefont {Zhong}, \citenamefont {Gu},
        \citenamefont {Fu}, \citenamefont {Ding},\ and\ \citenamefont {Gao}}]{re25}%
    \BibitemOpen
    \bibfield  {author} {\bibinfo {author} {\bibfnamefont {D.}~\bibnamefont
            {Wang}}, \bibinfo {author} {\bibfnamefont {L.}~\bibnamefont {Kong}}, \bibinfo
        {author} {\bibfnamefont {P.}~\bibnamefont {Fan}}, \bibinfo {author}
        {\bibfnamefont {H.}~\bibnamefont {Chen}}, \bibinfo {author} {\bibfnamefont
            {S.}~\bibnamefont {Zhu}}, \bibinfo {author} {\bibfnamefont {W.}~\bibnamefont
            {Liu}}, \bibinfo {author} {\bibfnamefont {L.}~\bibnamefont {Cao}}, \bibinfo
        {author} {\bibfnamefont {Y.}~\bibnamefont {Sun}}, \bibinfo {author}
        {\bibfnamefont {S.}~\bibnamefont {Du}}, \bibinfo {author} {\bibfnamefont
            {J.}~\bibnamefont {Schneeloch}}, \bibinfo {author} {\bibfnamefont
            {R.}~\bibnamefont {Zhong}}, \bibinfo {author} {\bibfnamefont
            {G.}~\bibnamefont {Gu}}, \bibinfo {author} {\bibfnamefont {L.}~\bibnamefont
            {Fu}}, \bibinfo {author} {\bibfnamefont {H.}~\bibnamefont {Ding}}, \ and\
        \bibinfo {author} {\bibfnamefont {H.-J.}\ \bibnamefont {Gao}},\ }\href
    {\doibase 10.1126/science.aao1797} {\bibfield  {journal} {\bibinfo  {journal}
            {Science}\ }\textbf {\bibinfo {volume} {362}},\ \bibinfo {pages} {333}
        (\bibinfo {year} {2018}{\natexlab{a}})}\BibitemShut {NoStop}%
    \bibitem [{\citenamefont {Liu}\ \emph {et~al.}(2018{\natexlab{a}})\citenamefont
        {Liu}, \citenamefont {Chen}, \citenamefont {Zhang}, \citenamefont {Peng},
        \citenamefont {Yan}, \citenamefont {Wen}, \citenamefont {Lou}, \citenamefont
        {Huang}, \citenamefont {Tian}, \citenamefont {Dong}, \citenamefont {Wang},
        \citenamefont {Bao}, \citenamefont {Wang}, \citenamefont {Yin}, \citenamefont
        {Zhao},\ and\ \citenamefont {Feng}}]{re26}%
    \BibitemOpen
    \bibfield  {author} {\bibinfo {author} {\bibfnamefont {Q.}~\bibnamefont
            {Liu}}, \bibinfo {author} {\bibfnamefont {C.}~\bibnamefont {Chen}}, \bibinfo
        {author} {\bibfnamefont {T.}~\bibnamefont {Zhang}}, \bibinfo {author}
        {\bibfnamefont {R.}~\bibnamefont {Peng}}, \bibinfo {author} {\bibfnamefont
            {Y.-J.}\ \bibnamefont {Yan}}, \bibinfo {author} {\bibfnamefont {C.-H.-P.}\
            \bibnamefont {Wen}}, \bibinfo {author} {\bibfnamefont {X.}~\bibnamefont
            {Lou}}, \bibinfo {author} {\bibfnamefont {Y.-L.}\ \bibnamefont {Huang}},
        \bibinfo {author} {\bibfnamefont {J.-P.}\ \bibnamefont {Tian}}, \bibinfo
        {author} {\bibfnamefont {X.-L.}\ \bibnamefont {Dong}}, \bibinfo {author}
        {\bibfnamefont {G.-W.}\ \bibnamefont {Wang}}, \bibinfo {author}
        {\bibfnamefont {W.-C.}\ \bibnamefont {Bao}}, \bibinfo {author} {\bibfnamefont
            {Q.-H.}\ \bibnamefont {Wang}}, \bibinfo {author} {\bibfnamefont {Z.-P.}\
            \bibnamefont {Yin}}, \bibinfo {author} {\bibfnamefont {Z.-X.}\ \bibnamefont
            {Zhao}}, \ and\ \bibinfo {author} {\bibfnamefont {D.-L.}\ \bibnamefont
            {Feng}},\ }\href {\doibase 10.1103/PhysRevX.8.041056} {\bibfield  {journal}
        {\bibinfo  {journal} {Phys. Rev. X}\ }\textbf {\bibinfo {volume} {8}},\
        \bibinfo {pages} {041056} (\bibinfo {year} {2018}{\natexlab{a}})}\BibitemShut
    {NoStop}%
    \bibitem [{\citenamefont {Benalcazar}\ \emph
        {et~al.}(2017{\natexlab{a}})\citenamefont {Benalcazar}, \citenamefont
        {Bernevig},\ and\ \citenamefont {Hughes}}]{benalcazar_quantized_2017}%
    \BibitemOpen
    \bibfield  {author} {\bibinfo {author} {\bibfnamefont {W.~A.}\ \bibnamefont
            {Benalcazar}}, \bibinfo {author} {\bibfnamefont {B.~A.}\ \bibnamefont
            {Bernevig}}, \ and\ \bibinfo {author} {\bibfnamefont {T.~L.}\ \bibnamefont
            {Hughes}},\ }\href {\doibase 10.1126/science.aah6442} {\bibfield  {journal}
        {\bibinfo  {journal} {Science}\ }\textbf {\bibinfo {volume} {357}},\ \bibinfo
        {pages} {61} (\bibinfo {year} {2017}{\natexlab{a}})}\BibitemShut {NoStop}%
    \bibitem [{\citenamefont {Benalcazar}\ \emph
        {et~al.}(2017{\natexlab{b}})\citenamefont {Benalcazar}, \citenamefont
        {Bernevig},\ and\ \citenamefont {Hughes}}]{benalcazar_electric_2017}%
    \BibitemOpen
    \bibfield  {author} {\bibinfo {author} {\bibfnamefont {W.~A.}\ \bibnamefont
            {Benalcazar}}, \bibinfo {author} {\bibfnamefont {B.~A.}\ \bibnamefont
            {Bernevig}}, \ and\ \bibinfo {author} {\bibfnamefont {T.~L.}\ \bibnamefont
            {Hughes}},\ }\href {\doibase 10.1103/PhysRevB.96.245115} {\bibfield
        {journal} {\bibinfo  {journal} {Phys. Rev. B}\ }\textbf {\bibinfo {volume}
            {96}},\ \bibinfo {pages} {245115} (\bibinfo {year}
        {2017}{\natexlab{b}})}\BibitemShut {NoStop}%
    \bibitem [{\citenamefont {Song}\ \emph {et~al.}(2017)\citenamefont {Song},
        \citenamefont {Fang},\ and\ \citenamefont {Fang}}]{song2017}%
    \BibitemOpen
    \bibfield  {author} {\bibinfo {author} {\bibfnamefont {Z.}~\bibnamefont
            {Song}}, \bibinfo {author} {\bibfnamefont {Z.}~\bibnamefont {Fang}}, \ and\
        \bibinfo {author} {\bibfnamefont {C.}~\bibnamefont {Fang}},\ }\href {\doibase
        10.1103/PhysRevLett.119.246402} {\bibfield  {journal} {\bibinfo  {journal}
            {Phys. Rev. Lett.}\ }\textbf {\bibinfo {volume} {119}},\ \bibinfo {pages}
        {246402} (\bibinfo {year} {2017})}\BibitemShut {NoStop}%
    \bibitem [{\citenamefont {Langbehn}\ \emph {et~al.}(2017)\citenamefont
        {Langbehn}, \citenamefont {Peng}, \citenamefont {Trifunovic}, \citenamefont
        {von Oppen},\ and\ \citenamefont {Brouwer}}]{re50}%
    \BibitemOpen
    \bibfield  {author} {\bibinfo {author} {\bibfnamefont {J.}~\bibnamefont
            {Langbehn}}, \bibinfo {author} {\bibfnamefont {Y.}~\bibnamefont {Peng}},
        \bibinfo {author} {\bibfnamefont {L.}~\bibnamefont {Trifunovic}}, \bibinfo
        {author} {\bibfnamefont {F.}~\bibnamefont {von Oppen}}, \ and\ \bibinfo
        {author} {\bibfnamefont {P.~W.}\ \bibnamefont {Brouwer}},\ }\href {\doibase
        10.1103/PhysRevLett.119.246401} {\bibfield  {journal} {\bibinfo  {journal}
            {Phys. Rev. Lett.}\ }\textbf {\bibinfo {volume} {119}},\ \bibinfo {pages}
        {246401} (\bibinfo {year} {2017})}\BibitemShut {NoStop}%
    \bibitem [{\citenamefont {Schindler}\ \emph {et~al.}(2018)\citenamefont
        {Schindler}, \citenamefont {Cook}, \citenamefont {Vergniory}, \citenamefont
        {Wang}, \citenamefont {Parkin}, \citenamefont {Bernevig},\ and\ \citenamefont
        {Neupert}}]{schindler_2018}%
    \BibitemOpen
    \bibfield  {author} {\bibinfo {author} {\bibfnamefont {F.}~\bibnamefont
            {Schindler}}, \bibinfo {author} {\bibfnamefont {A.~M.}\ \bibnamefont {Cook}},
        \bibinfo {author} {\bibfnamefont {M.~G.}\ \bibnamefont {Vergniory}}, \bibinfo
        {author} {\bibfnamefont {Z.}~\bibnamefont {Wang}}, \bibinfo {author}
        {\bibfnamefont {S.~S.~P.}\ \bibnamefont {Parkin}}, \bibinfo {author}
        {\bibfnamefont {B.~A.}\ \bibnamefont {Bernevig}}, \ and\ \bibinfo {author}
        {\bibfnamefont {T.}~\bibnamefont {Neupert}},\ }\href {\doibase
        10.1126/sciadv.aat0346} {\bibfield  {journal} {\bibinfo  {journal} {Sci.
                Adv.}\ }\textbf {\bibinfo {volume} {4}},\ \bibinfo {pages} {eaat0346}
        (\bibinfo {year} {2018})}\BibitemShut {NoStop}%
    \bibitem [{\citenamefont {Geier}\ \emph {et~al.}(2018)\citenamefont {Geier},
        \citenamefont {Trifunovic}, \citenamefont {Hoskam},\ and\ \citenamefont
        {Brouwer}}]{re51}%
    \BibitemOpen
    \bibfield  {author} {\bibinfo {author} {\bibfnamefont {M.}~\bibnamefont
            {Geier}}, \bibinfo {author} {\bibfnamefont {L.}~\bibnamefont {Trifunovic}},
        \bibinfo {author} {\bibfnamefont {M.}~\bibnamefont {Hoskam}}, \ and\ \bibinfo
        {author} {\bibfnamefont {P.~W.}\ \bibnamefont {Brouwer}},\ }\href {\doibase
        10.1103/PhysRevB.97.205135} {\bibfield  {journal} {\bibinfo  {journal} {Phys.
                Rev. B}\ }\textbf {\bibinfo {volume} {97}},\ \bibinfo {pages} {205135}
        (\bibinfo {year} {2018})}\BibitemShut {NoStop}%
    \bibitem [{\citenamefont {Yan}\ \emph {et~al.}(2018)\citenamefont {Yan},
        \citenamefont {Song},\ and\ \citenamefont {Wang}}]{yandw}%
    \BibitemOpen
    \bibfield  {author} {\bibinfo {author} {\bibfnamefont {Z.}~\bibnamefont
            {Yan}}, \bibinfo {author} {\bibfnamefont {F.}~\bibnamefont {Song}}, \ and\
        \bibinfo {author} {\bibfnamefont {Z.}~\bibnamefont {Wang}},\ }\href {\doibase
        10.1103/PhysRevLett.121.096803} {\bibfield  {journal} {\bibinfo  {journal}
            {Phys. Rev. Lett.}\ }\textbf {\bibinfo {volume} {121}},\ \bibinfo {pages}
        {096803} (\bibinfo {year} {2018})}\BibitemShut {NoStop}%
    \bibitem [{\citenamefont {Wang}\ \emph
        {et~al.}(2018{\natexlab{b}})\citenamefont {Wang}, \citenamefont {Liu},
        \citenamefont {Lu},\ and\ \citenamefont {Zhang}}]{wangspm}%
    \BibitemOpen
    \bibfield  {author} {\bibinfo {author} {\bibfnamefont {Q.}~\bibnamefont
            {Wang}}, \bibinfo {author} {\bibfnamefont {C.-C.}\ \bibnamefont {Liu}},
        \bibinfo {author} {\bibfnamefont {Y.-M.}\ \bibnamefont {Lu}}, \ and\ \bibinfo
        {author} {\bibfnamefont {F.}~\bibnamefont {Zhang}},\ }\href {\doibase
        10.1103/PhysRevLett.121.186801} {\bibfield  {journal} {\bibinfo  {journal}
            {Phys. Rev. Lett.}\ }\textbf {\bibinfo {volume} {121}},\ \bibinfo {pages}
        {186801} (\bibinfo {year} {2018}{\natexlab{b}})}\BibitemShut {NoStop}%
    \bibitem [{\citenamefont {Zhu}(2018)}]{zhuso}%
    \BibitemOpen
    \bibfield  {author} {\bibinfo {author} {\bibfnamefont {X.}~\bibnamefont
            {Zhu}},\ }\href {\doibase 10.1103/PhysRevB.97.205134} {\bibfield  {journal}
        {\bibinfo  {journal} {Phys. Rev. B}\ }\textbf {\bibinfo {volume} {97}},\
        \bibinfo {pages} {205134} (\bibinfo {year} {2018})}\BibitemShut {NoStop}%
    \bibitem [{\citenamefont {Hsu}\ \emph {et~al.}(2018)\citenamefont {Hsu},
        \citenamefont {Stano}, \citenamefont {Klinovaja},\ and\ \citenamefont
        {Loss}}]{re49}%
    \BibitemOpen
    \bibfield  {author} {\bibinfo {author} {\bibfnamefont {C.-H.}\ \bibnamefont
            {Hsu}}, \bibinfo {author} {\bibfnamefont {P.}~\bibnamefont {Stano}}, \bibinfo
        {author} {\bibfnamefont {J.}~\bibnamefont {Klinovaja}}, \ and\ \bibinfo
        {author} {\bibfnamefont {D.}~\bibnamefont {Loss}},\ }\href {\doibase
        10.1103/PhysRevLett.121.196801} {\bibfield  {journal} {\bibinfo  {journal}
            {Phys. Rev. Lett.}\ }\textbf {\bibinfo {volume} {121}},\ \bibinfo {pages}
        {196801} (\bibinfo {year} {2018})}\BibitemShut {NoStop}%
    \bibitem [{\citenamefont {Khalaf}(2018)}]{re39}%
    \BibitemOpen
    \bibfield  {author} {\bibinfo {author} {\bibfnamefont {E.}~\bibnamefont
            {Khalaf}},\ }\href {\doibase 10.1103/PhysRevB.97.205136} {\bibfield
        {journal} {\bibinfo  {journal} {Phys. Rev. B}\ }\textbf {\bibinfo {volume}
            {97}},\ \bibinfo {pages} {205136} (\bibinfo {year} {2018})}\BibitemShut
    {NoStop}%
    \bibitem [{\citenamefont {Liu}\ \emph {et~al.}(2018{\natexlab{b}})\citenamefont
        {Liu}, \citenamefont {He},\ and\ \citenamefont {Nori}}]{liu_majorana_2018}%
    \BibitemOpen
    \bibfield  {author} {\bibinfo {author} {\bibfnamefont {T.}~\bibnamefont
            {Liu}}, \bibinfo {author} {\bibfnamefont {J.~J.}\ \bibnamefont {He}}, \ and\
        \bibinfo {author} {\bibfnamefont {F.}~\bibnamefont {Nori}},\ }\href {\doibase
        10.1103/PhysRevB.98.245413} {\bibfield  {journal} {\bibinfo  {journal} {Phys.
                Rev. B}\ }\textbf {\bibinfo {volume} {98}},\ \bibinfo {pages} {245413}
        (\bibinfo {year} {2018}{\natexlab{b}})}\BibitemShut {NoStop}%
    \bibitem [{\citenamefont {Wang}\ \emph
        {et~al.}(2018{\natexlab{c}})\citenamefont {Wang}, \citenamefont {Lin},\ and\
        \citenamefont {Hughes}}]{yxwangprb}%
    \BibitemOpen
    \bibfield  {author} {\bibinfo {author} {\bibfnamefont {Y.}~\bibnamefont
            {Wang}}, \bibinfo {author} {\bibfnamefont {M.}~\bibnamefont {Lin}}, \ and\
        \bibinfo {author} {\bibfnamefont {T.~L.}\ \bibnamefont {Hughes}},\ }\href
    {\doibase 10.1103/PhysRevB.98.165144} {\bibfield  {journal} {\bibinfo
            {journal} {Phys. Rev. B}\ }\textbf {\bibinfo {volume} {98}},\ \bibinfo
        {pages} {165144} (\bibinfo {year} {2018}{\natexlab{c}})}\BibitemShut
    {NoStop}%
    \bibitem [{\citenamefont {Volpez}\ \emph {et~al.}(2019)\citenamefont {Volpez},
        \citenamefont {Loss},\ and\ \citenamefont {Klinovaja}}]{re53}%
    \BibitemOpen
    \bibfield  {author} {\bibinfo {author} {\bibfnamefont {Y.}~\bibnamefont
            {Volpez}}, \bibinfo {author} {\bibfnamefont {D.}~\bibnamefont {Loss}}, \ and\
        \bibinfo {author} {\bibfnamefont {J.}~\bibnamefont {Klinovaja}},\ }\href
    {\doibase 10.1103/PhysRevLett.122.126402} {\bibfield  {journal} {\bibinfo
            {journal} {Phys. Rev. Lett.}\ }\textbf {\bibinfo {volume} {122}},\ \bibinfo
        {pages} {126402} (\bibinfo {year} {2019})}\BibitemShut {NoStop}%
    \bibitem [{\citenamefont {Zhang}\ \emph
        {et~al.}(2019{\natexlab{a}})\citenamefont {Zhang}, \citenamefont {Cole},\
        and\ \citenamefont {Das~Sarma}}]{RXzhang-iron}%
    \BibitemOpen
    \bibfield  {author} {\bibinfo {author} {\bibfnamefont {R.-X.}\ \bibnamefont
            {Zhang}}, \bibinfo {author} {\bibfnamefont {W.~S.}\ \bibnamefont {Cole}}, \
        and\ \bibinfo {author} {\bibfnamefont {S.}~\bibnamefont {Das~Sarma}},\ }\href
    {\doibase 10.1103/PhysRevLett.122.187001} {\bibfield  {journal} {\bibinfo
            {journal} {Phys. Rev. Lett.}\ }\textbf {\bibinfo {volume} {122}},\ \bibinfo
        {pages} {187001} (\bibinfo {year} {2019}{\natexlab{a}})}\BibitemShut
    {NoStop}%
    \bibitem [{\citenamefont {Zhu}(2019)}]{zhumix}%
    \BibitemOpen
    \bibfield  {author} {\bibinfo {author} {\bibfnamefont {X.}~\bibnamefont
            {Zhu}},\ }\href {\doibase 10.1103/PhysRevLett.122.236401} {\bibfield
        {journal} {\bibinfo  {journal} {Phys. Rev. Lett.}\ }\textbf {\bibinfo
            {volume} {122}},\ \bibinfo {pages} {236401} (\bibinfo {year}
        {2019})}\BibitemShut {NoStop}%
    \bibitem [{\citenamefont {Yan}(2019)}]{yanodd}%
    \BibitemOpen
    \bibfield  {author} {\bibinfo {author} {\bibfnamefont {Z.}~\bibnamefont
            {Yan}},\ }\href {\doibase 10.1103/PhysRevLett.123.177001} {\bibfield
        {journal} {\bibinfo  {journal} {Phys. Rev. Lett.}\ }\textbf {\bibinfo
            {volume} {123}},\ \bibinfo {pages} {177001} (\bibinfo {year}
        {2019})}\BibitemShut {NoStop}%
    \bibitem [{\citenamefont {Zeng}\ \emph {et~al.}(2019)\citenamefont {Zeng},
        \citenamefont {Stanescu}, \citenamefont {Zhang}, \citenamefont {Scarola},\
        and\ \citenamefont {Tewari}}]{zccho}%
    \BibitemOpen
    \bibfield  {author} {\bibinfo {author} {\bibfnamefont {C.}~\bibnamefont
            {Zeng}}, \bibinfo {author} {\bibfnamefont {T.~D.}\ \bibnamefont {Stanescu}},
        \bibinfo {author} {\bibfnamefont {C.}~\bibnamefont {Zhang}}, \bibinfo
        {author} {\bibfnamefont {V.~W.}\ \bibnamefont {Scarola}}, \ and\ \bibinfo
        {author} {\bibfnamefont {S.}~\bibnamefont {Tewari}},\ }\href {\doibase
        10.1103/PhysRevLett.123.060402} {\bibfield  {journal} {\bibinfo  {journal}
            {Phys. Rev. Lett.}\ }\textbf {\bibinfo {volume} {123}},\ \bibinfo {pages}
        {060402} (\bibinfo {year} {2019})}\BibitemShut {NoStop}%
    \bibitem [{\citenamefont {Zhang}\ \emph
        {et~al.}(2019{\natexlab{b}})\citenamefont {Zhang}, \citenamefont {Cole},
        \citenamefont {Wu},\ and\ \citenamefont {Das~Sarma}}]{re42}%
    \BibitemOpen
    \bibfield  {author} {\bibinfo {author} {\bibfnamefont {R.-X.}\ \bibnamefont
            {Zhang}}, \bibinfo {author} {\bibfnamefont {W.~S.}\ \bibnamefont {Cole}},
        \bibinfo {author} {\bibfnamefont {X.}~\bibnamefont {Wu}}, \ and\ \bibinfo
        {author} {\bibfnamefont {S.}~\bibnamefont {Das~Sarma}},\ }\href {\doibase
        10.1103/PhysRevLett.123.167001} {\bibfield  {journal} {\bibinfo  {journal}
            {Phys. Rev. Lett.}\ }\textbf {\bibinfo {volume} {123}},\ \bibinfo {pages}
        {167001} (\bibinfo {year} {2019}{\natexlab{b}})}\BibitemShut {NoStop}%
    \bibitem [{\citenamefont {Peng}\ and\ \citenamefont {Xu}(2019)}]{pyati}%
    \BibitemOpen
    \bibfield  {author} {\bibinfo {author} {\bibfnamefont {Y.}~\bibnamefont
            {Peng}}\ and\ \bibinfo {author} {\bibfnamefont {Y.}~\bibnamefont {Xu}},\
    }\href {\doibase 10.1103/PhysRevB.99.195431} {\bibfield  {journal} {\bibinfo
            {journal} {Phys. Rev. B}\ }\textbf {\bibinfo {volume} {99}},\ \bibinfo
        {pages} {195431} (\bibinfo {year} {2019})}\BibitemShut {NoStop}%
    \bibitem [{\citenamefont {Pan}\ \emph {et~al.}(2019)\citenamefont {Pan},
        \citenamefont {Yang}, \citenamefont {Chen}, \citenamefont {Xu}, \citenamefont
        {Liu},\ and\ \citenamefont {Liu}}]{cxliu-htsc}%
    \BibitemOpen
    \bibfield  {author} {\bibinfo {author} {\bibfnamefont {X.-H.}\ \bibnamefont
            {Pan}}, \bibinfo {author} {\bibfnamefont {K.-J.}\ \bibnamefont {Yang}},
        \bibinfo {author} {\bibfnamefont {L.}~\bibnamefont {Chen}}, \bibinfo {author}
        {\bibfnamefont {G.}~\bibnamefont {Xu}}, \bibinfo {author} {\bibfnamefont
            {C.-X.}\ \bibnamefont {Liu}}, \ and\ \bibinfo {author} {\bibfnamefont
            {X.}~\bibnamefont {Liu}},\ }\href {\doibase 10.1103/PhysRevLett.123.156801}
    {\bibfield  {journal} {\bibinfo  {journal} {Phys. Rev. Lett.}\ }\textbf
        {\bibinfo {volume} {123}},\ \bibinfo {pages} {156801} (\bibinfo {year}
        {2019})}\BibitemShut {NoStop}%
    \bibitem [{\citenamefont {Franca}\ \emph {et~al.}(2019)\citenamefont {Franca},
        \citenamefont {Efremov},\ and\ \citenamefont {Fulga}}]{Fulga_prb}%
    \BibitemOpen
    \bibfield  {author} {\bibinfo {author} {\bibfnamefont {S.}~\bibnamefont
            {Franca}}, \bibinfo {author} {\bibfnamefont {D.~V.}\ \bibnamefont {Efremov}},
        \ and\ \bibinfo {author} {\bibfnamefont {I.~C.}\ \bibnamefont {Fulga}},\
    }\href {\doibase 10.1103/PhysRevB.100.075415} {\bibfield  {journal} {\bibinfo
            {journal} {Phys. Rev. B}\ }\textbf {\bibinfo {volume} {100}},\ \bibinfo
        {pages} {075415} (\bibinfo {year} {2019})}\BibitemShut {NoStop}%
    \bibitem [{\citenamefont {Trifunovic}\ and\ \citenamefont
        {Brouwer}(2019)}]{Piet_prx}%
    \BibitemOpen
    \bibfield  {author} {\bibinfo {author} {\bibfnamefont {L.}~\bibnamefont
            {Trifunovic}}\ and\ \bibinfo {author} {\bibfnamefont {P.~W.}\ \bibnamefont
            {Brouwer}},\ }\href {\doibase 10.1103/PhysRevX.9.011012} {\bibfield
        {journal} {\bibinfo  {journal} {Phys. Rev. X}\ }\textbf {\bibinfo {volume}
            {9}},\ \bibinfo {pages} {011012} (\bibinfo {year} {2019})}\BibitemShut
    {NoStop}%
    \bibitem [{\citenamefont {Gray}\ \emph {et~al.}(2019)\citenamefont {Gray},
        \citenamefont {Freudenstein}, \citenamefont {Zhao}, \citenamefont
        {\'OConnor}, \citenamefont {Jenkins}, \citenamefont {Kumar}, \citenamefont
        {Hoek}, \citenamefont {Kopec}, \citenamefont {Huh}, \citenamefont
        {Taniguchi}, \citenamefont {Watanabe}, \citenamefont {Zhong}, \citenamefont
        {Kim}, \citenamefont {Gu},\ and\ \citenamefont {Burch}}]{re68}%
    \BibitemOpen
    \bibfield  {author} {\bibinfo {author} {\bibfnamefont {M.~J.}\ \bibnamefont
            {Gray}}, \bibinfo {author} {\bibfnamefont {J.}~\bibnamefont {Freudenstein}},
        \bibinfo {author} {\bibfnamefont {S.~Y.~F.}\ \bibnamefont {Zhao}}, \bibinfo
        {author} {\bibfnamefont {R.}~\bibnamefont {\'OConnor}}, \bibinfo {author}
        {\bibfnamefont {S.}~\bibnamefont {Jenkins}}, \bibinfo {author} {\bibfnamefont
            {N.}~\bibnamefont {Kumar}}, \bibinfo {author} {\bibfnamefont
            {M.}~\bibnamefont {Hoek}}, \bibinfo {author} {\bibfnamefont {A.}~\bibnamefont
            {Kopec}}, \bibinfo {author} {\bibfnamefont {S.}~\bibnamefont {Huh}}, \bibinfo
        {author} {\bibfnamefont {T.}~\bibnamefont {Taniguchi}}, \bibinfo {author}
        {\bibfnamefont {K.}~\bibnamefont {Watanabe}}, \bibinfo {author}
        {\bibfnamefont {R.}~\bibnamefont {Zhong}}, \bibinfo {author} {\bibfnamefont
            {C.}~\bibnamefont {Kim}}, \bibinfo {author} {\bibfnamefont {G.~D.}\
            \bibnamefont {Gu}}, \ and\ \bibinfo {author} {\bibfnamefont {K.~S.}\
            \bibnamefont {Burch}},\ }\href {\doibase 10.1021/acs.nanolett.9b00844}
    {\bibfield  {journal} {\bibinfo  {journal} {Nano Letters}\ }\textbf {\bibinfo
            {volume} {19}},\ \bibinfo {pages} {4890} (\bibinfo {year}
        {2019})}\BibitemShut {NoStop}%
    \bibitem [{\citenamefont {Wu}\ \emph {et~al.}(2020{\natexlab{a}})\citenamefont
        {Wu}, \citenamefont {Hou}, \citenamefont {Li}, \citenamefont {Luo},
        \citenamefont {Shi},\ and\ \citenamefont {Zhang}}]{re43}%
    \BibitemOpen
    \bibfield  {author} {\bibinfo {author} {\bibfnamefont {Y.-J.}\ \bibnamefont
            {Wu}}, \bibinfo {author} {\bibfnamefont {J.}~\bibnamefont {Hou}}, \bibinfo
        {author} {\bibfnamefont {Y.-M.}\ \bibnamefont {Li}}, \bibinfo {author}
        {\bibfnamefont {X.-W.}\ \bibnamefont {Luo}}, \bibinfo {author} {\bibfnamefont
            {X.}~\bibnamefont {Shi}}, \ and\ \bibinfo {author} {\bibfnamefont
            {C.}~\bibnamefont {Zhang}},\ }\href {\doibase 10.1103/PhysRevLett.124.227001}
    {\bibfield  {journal} {\bibinfo  {journal} {Phys. Rev. Lett.}\ }\textbf
        {\bibinfo {volume} {124}},\ \bibinfo {pages} {227001} (\bibinfo {year}
        {2020}{\natexlab{a}})}\BibitemShut {NoStop}%
    \bibitem [{\citenamefont {Ahn}\ and\ \citenamefont {Yang}(2020)}]{bjyangtsc}%
    \BibitemOpen
    \bibfield  {author} {\bibinfo {author} {\bibfnamefont {J.}~\bibnamefont
            {Ahn}}\ and\ \bibinfo {author} {\bibfnamefont {B.-J.}\ \bibnamefont {Yang}},\
    }\href {\doibase 10.1103/PhysRevResearch.2.012060} {\bibfield  {journal}
        {\bibinfo  {journal} {Phys. Rev. Research}\ }\textbf {\bibinfo {volume}
            {2}},\ \bibinfo {pages} {012060} (\bibinfo {year} {2020})}\BibitemShut
    {NoStop}%
    \bibitem [{\citenamefont {Wu}\ \emph {et~al.}(2020{\natexlab{b}})\citenamefont
        {Wu}, \citenamefont {Benalcazar}, \citenamefont {Li}, \citenamefont
        {Thomale}, \citenamefont {Liu},\ and\ \citenamefont {Hu}}]{re35}%
    \BibitemOpen
    \bibfield  {author} {\bibinfo {author} {\bibfnamefont {X.}~\bibnamefont
            {Wu}}, \bibinfo {author} {\bibfnamefont {W.~A.}\ \bibnamefont {Benalcazar}},
        \bibinfo {author} {\bibfnamefont {Y.}~\bibnamefont {Li}}, \bibinfo {author}
        {\bibfnamefont {R.}~\bibnamefont {Thomale}}, \bibinfo {author} {\bibfnamefont
            {C.-X.}\ \bibnamefont {Liu}}, \ and\ \bibinfo {author} {\bibfnamefont
            {J.}~\bibnamefont {Hu}},\ }\href {\doibase 10.1103/PhysRevX.10.041014}
    {\bibfield  {journal} {\bibinfo  {journal} {Phys. Rev. X}\ }\textbf {\bibinfo
            {volume} {10}},\ \bibinfo {pages} {041014} (\bibinfo {year}
        {2020}{\natexlab{b}})}\BibitemShut {NoStop}%
    \bibitem [{\citenamefont {Luo}\ \emph {et~al.}(2021)\citenamefont {Luo},
        \citenamefont {Pan},\ and\ \citenamefont {Liu}}]{re41}%
    \BibitemOpen
    \bibfield  {author} {\bibinfo {author} {\bibfnamefont {X.-J.}\ \bibnamefont
            {Luo}}, \bibinfo {author} {\bibfnamefont {X.-H.}\ \bibnamefont {Pan}}, \ and\
        \bibinfo {author} {\bibfnamefont {X.}~\bibnamefont {Liu}},\ }\href {\doibase
        10.1103/PhysRevB.104.104510} {\bibfield  {journal} {\bibinfo  {journal}
            {Phys. Rev. B}\ }\textbf {\bibinfo {volume} {104}},\ \bibinfo {pages}
        {104510} (\bibinfo {year} {2021})}\BibitemShut {NoStop}%
    \bibitem [{\citenamefont {Zhang}\ \emph {et~al.}(2020)\citenamefont {Zhang},
        \citenamefont {Rui}, \citenamefont {Calzona}, \citenamefont {Choi},
        \citenamefont {Schnyder},\ and\ \citenamefont {Trauzettel}}]{sbzhangtsc}%
    \BibitemOpen
    \bibfield  {author} {\bibinfo {author} {\bibfnamefont {S.-B.}\ \bibnamefont
            {Zhang}}, \bibinfo {author} {\bibfnamefont {W.~B.}\ \bibnamefont {Rui}},
        \bibinfo {author} {\bibfnamefont {A.}~\bibnamefont {Calzona}}, \bibinfo
        {author} {\bibfnamefont {S.-J.}\ \bibnamefont {Choi}}, \bibinfo {author}
        {\bibfnamefont {A.~P.}\ \bibnamefont {Schnyder}}, \ and\ \bibinfo {author}
        {\bibfnamefont {B.}~\bibnamefont {Trauzettel}},\ }\href {\doibase
        10.1103/PhysRevResearch.2.043025} {\bibfield  {journal} {\bibinfo  {journal}
            {Phys. Rev. Research}\ }\textbf {\bibinfo {volume} {2}},\ \bibinfo {pages}
        {043025} (\bibinfo {year} {2020})}\BibitemShut {NoStop}%
    \bibitem [{\citenamefont {Kheirkhah}\ \emph {et~al.}(2020)\citenamefont
        {Kheirkhah}, \citenamefont {Yan}, \citenamefont {Nagai},\ and\ \citenamefont
        {Marsiglio}}]{khhubbrd}%
    \BibitemOpen
    \bibfield  {author} {\bibinfo {author} {\bibfnamefont {M.}~\bibnamefont
            {Kheirkhah}}, \bibinfo {author} {\bibfnamefont {Z.}~\bibnamefont {Yan}},
        \bibinfo {author} {\bibfnamefont {Y.}~\bibnamefont {Nagai}}, \ and\ \bibinfo
        {author} {\bibfnamefont {F.}~\bibnamefont {Marsiglio}},\ }\href {\doibase
        10.1103/PhysRevLett.125.017001} {\bibfield  {journal} {\bibinfo  {journal}
            {Phys. Rev. Lett.}\ }\textbf {\bibinfo {volume} {125}},\ \bibinfo {pages}
        {017001} (\bibinfo {year} {2020})}\BibitemShut {NoStop}%
    \bibitem [{\citenamefont {Ghorashi}\ \emph {et~al.}(2020)\citenamefont
        {Ghorashi}, \citenamefont {Hughes},\ and\ \citenamefont {Rossi}}]{ab-htsc}%
    \BibitemOpen
    \bibfield  {author} {\bibinfo {author} {\bibfnamefont {S.~A.~A.}\
            \bibnamefont {Ghorashi}}, \bibinfo {author} {\bibfnamefont {T.~L.}\
            \bibnamefont {Hughes}}, \ and\ \bibinfo {author} {\bibfnamefont
            {E.}~\bibnamefont {Rossi}},\ }\href {\doibase 10.1103/PhysRevLett.125.037001}
    {\bibfield  {journal} {\bibinfo  {journal} {Phys. Rev. Lett.}\ }\textbf
        {\bibinfo {volume} {125}},\ \bibinfo {pages} {037001} (\bibinfo {year}
        {2020})}\BibitemShut {NoStop}%
    \bibitem [{\citenamefont {Ghosh}\ \emph {et~al.}(2021)\citenamefont {Ghosh},
        \citenamefont {Nag},\ and\ \citenamefont {Saha}}]{saha-Floquet}%
    \BibitemOpen
    \bibfield  {author} {\bibinfo {author} {\bibfnamefont {A.~K.}\ \bibnamefont
            {Ghosh}}, \bibinfo {author} {\bibfnamefont {T.}~\bibnamefont {Nag}}, \ and\
        \bibinfo {author} {\bibfnamefont {A.}~\bibnamefont {Saha}},\ }\href {\doibase
        10.1103/PhysRevB.103.085413} {\bibfield  {journal} {\bibinfo  {journal}
            {Phys. Rev. B}\ }\textbf {\bibinfo {volume} {103}},\ \bibinfo {pages}
        {085413} (\bibinfo {year} {2021})}\BibitemShut {NoStop}%
    \bibitem [{\citenamefont {Fu}\ \emph {et~al.}(2021)\citenamefont {Fu},
        \citenamefont {Hu}, \citenamefont {Li}, \citenamefont {Li},\ and\
        \citenamefont {Shen}}]{Shenchiral}%
    \BibitemOpen
    \bibfield  {author} {\bibinfo {author} {\bibfnamefont {B.}~\bibnamefont
            {Fu}}, \bibinfo {author} {\bibfnamefont {Z.-A.}\ \bibnamefont {Hu}}, \bibinfo
        {author} {\bibfnamefont {C.-A.}\ \bibnamefont {Li}}, \bibinfo {author}
        {\bibfnamefont {J.}~\bibnamefont {Li}}, \ and\ \bibinfo {author}
        {\bibfnamefont {S.-Q.}\ \bibnamefont {Shen}},\ }\href {\doibase
        10.1103/PhysRevB.103.L180504} {\bibfield  {journal} {\bibinfo  {journal}
            {Phys. Rev. B}\ }\textbf {\bibinfo {volume} {103}},\ \bibinfo {pages}
        {L180504} (\bibinfo {year} {2021})}\BibitemShut {NoStop}%
    \bibitem [{\citenamefont {Qin}\ \emph {et~al.}(2022)\citenamefont {Qin},
        \citenamefont {Fang}, \citenamefont {Zhang},\ and\ \citenamefont
        {Hu}}]{qin_topological_2022}%
    \BibitemOpen
    \bibfield  {author} {\bibinfo {author} {\bibfnamefont {S.}~\bibnamefont
            {Qin}}, \bibinfo {author} {\bibfnamefont {C.}~\bibnamefont {Fang}}, \bibinfo
        {author} {\bibfnamefont {F.-C.}\ \bibnamefont {Zhang}}, \ and\ \bibinfo
        {author} {\bibfnamefont {J.}~\bibnamefont {Hu}},\ }\href {\doibase
        10.1103/PhysRevX.12.011030} {\bibfield  {journal} {\bibinfo  {journal} {Phys.
                Rev. X}\ }\textbf {\bibinfo {volume} {12}},\ \bibinfo {pages} {011030}
        (\bibinfo {year} {2022})}\BibitemShut {NoStop}%
    \bibitem [{\citenamefont {Tan}\ \emph {et~al.}(2022)\citenamefont {Tan},
        \citenamefont {Huang},\ and\ \citenamefont {Liu}}]{xjLiuprb}%
    \BibitemOpen
    \bibfield  {author} {\bibinfo {author} {\bibfnamefont {Y.}~\bibnamefont
            {Tan}}, \bibinfo {author} {\bibfnamefont {Z.-H.}\ \bibnamefont {Huang}}, \
        and\ \bibinfo {author} {\bibfnamefont {X.-J.}\ \bibnamefont {Liu}},\ }\href
    {\doibase 10.1103/PhysRevB.105.L041105} {\bibfield  {journal} {\bibinfo
            {journal} {Phys. Rev. B}\ }\textbf {\bibinfo {volume} {105}},\ \bibinfo
        {pages} {L041105} (\bibinfo {year} {2022})}\BibitemShut {NoStop}%
    \bibitem [{\citenamefont {Wu}\ \emph {et~al.}(2022)\citenamefont {Wu},
        \citenamefont {Liu}, \citenamefont {Thomale},\ and\ \citenamefont
        {Liu}}]{xxwu2}%
    \BibitemOpen
    \bibfield  {author} {\bibinfo {author} {\bibfnamefont {X.}~\bibnamefont
            {Wu}}, \bibinfo {author} {\bibfnamefont {X.}~\bibnamefont {Liu}}, \bibinfo
        {author} {\bibfnamefont {R.}~\bibnamefont {Thomale}}, \ and\ \bibinfo
        {author} {\bibfnamefont {C.-X.}\ \bibnamefont {Liu}},\ }\href {\doibase
        10.1093/nsr/nwab087} {\bibfield  {journal} {\bibinfo  {journal} {Natl. Sci.
                Rev.}\ }\textbf {\bibinfo {volume} {9}},\ \bibinfo {pages} {nwab087}
        (\bibinfo {year} {2022})}\BibitemShut {NoStop}%
    \bibitem [{\citenamefont {Cao}\ \emph {et~al.}(2018{\natexlab{a}})\citenamefont
        {Cao}, \citenamefont {Fatemi}, \citenamefont {Demir}, \citenamefont {Fang},
        \citenamefont {Tomarken}, \citenamefont {Luo}, \citenamefont
        {Sanchez-Yamagishi}, \citenamefont {Watanabe}, \citenamefont {Taniguchi},
        \citenamefont {Kaxiras}, \citenamefont {Ashoori},\ and\ \citenamefont
        {Jarillo-Herrero}}]{tbg1}%
    \BibitemOpen
    \bibfield  {author} {\bibinfo {author} {\bibfnamefont {Y.}~\bibnamefont
            {Cao}}, \bibinfo {author} {\bibfnamefont {V.}~\bibnamefont {Fatemi}},
        \bibinfo {author} {\bibfnamefont {A.}~\bibnamefont {Demir}}, \bibinfo
        {author} {\bibfnamefont {S.}~\bibnamefont {Fang}}, \bibinfo {author}
        {\bibfnamefont {S.~L.}\ \bibnamefont {Tomarken}}, \bibinfo {author}
        {\bibfnamefont {J.~Y.}\ \bibnamefont {Luo}}, \bibinfo {author} {\bibfnamefont
            {J.~D.}\ \bibnamefont {Sanchez-Yamagishi}}, \bibinfo {author} {\bibfnamefont
            {K.}~\bibnamefont {Watanabe}}, \bibinfo {author} {\bibfnamefont
            {T.}~\bibnamefont {Taniguchi}}, \bibinfo {author} {\bibfnamefont
            {E.}~\bibnamefont {Kaxiras}}, \bibinfo {author} {\bibfnamefont {R.~C.}\
            \bibnamefont {Ashoori}}, \ and\ \bibinfo {author} {\bibfnamefont
            {P.}~\bibnamefont {Jarillo-Herrero}},\ }\href {\doibase 10.1038/nature26154}
    {\bibfield  {journal} {\bibinfo  {journal} {Nature}\ }\textbf {\bibinfo
            {volume} {556}},\ \bibinfo {pages} {80} (\bibinfo {year}
        {2018}{\natexlab{a}})}\BibitemShut {NoStop}%
    \bibitem [{\citenamefont {Cao}\ \emph {et~al.}(2018{\natexlab{b}})\citenamefont
        {Cao}, \citenamefont {Fatemi}, \citenamefont {Fang}, \citenamefont
        {Watanabe}, \citenamefont {Taniguchi}, \citenamefont {Kaxiras},\ and\
        \citenamefont {Jarillo-Herrero}}]{tbg2}%
    \BibitemOpen
    \bibfield  {author} {\bibinfo {author} {\bibfnamefont {Y.}~\bibnamefont
            {Cao}}, \bibinfo {author} {\bibfnamefont {V.}~\bibnamefont {Fatemi}},
        \bibinfo {author} {\bibfnamefont {S.}~\bibnamefont {Fang}}, \bibinfo {author}
        {\bibfnamefont {K.}~\bibnamefont {Watanabe}}, \bibinfo {author}
        {\bibfnamefont {T.}~\bibnamefont {Taniguchi}}, \bibinfo {author}
        {\bibfnamefont {E.}~\bibnamefont {Kaxiras}}, \ and\ \bibinfo {author}
        {\bibfnamefont {P.}~\bibnamefont {Jarillo-Herrero}},\ }\href {\doibase
        10.1038/nature26160} {\bibfield  {journal} {\bibinfo  {journal} {Nature}\
        }\textbf {\bibinfo {volume} {556}},\ \bibinfo {pages} {43} (\bibinfo {year}
        {2018}{\natexlab{b}})}\BibitemShut {NoStop}%
    \bibitem [{\citenamefont {Yu}\ \emph {et~al.}(2019)\citenamefont {Yu},
        \citenamefont {Ma}, \citenamefont {Cai}, \citenamefont {Zhong}, \citenamefont
        {Ye}, \citenamefont {Shen}, \citenamefont {Gu}, \citenamefont {Chen},\ and\
        \citenamefont {Zhang}}]{scmo}%
    \BibitemOpen
    \bibfield  {author} {\bibinfo {author} {\bibfnamefont {Y.}~\bibnamefont
            {Yu}}, \bibinfo {author} {\bibfnamefont {L.}~\bibnamefont {Ma}}, \bibinfo
        {author} {\bibfnamefont {P.}~\bibnamefont {Cai}}, \bibinfo {author}
        {\bibfnamefont {R.}~\bibnamefont {Zhong}}, \bibinfo {author} {\bibfnamefont
            {C.}~\bibnamefont {Ye}}, \bibinfo {author} {\bibfnamefont {J.}~\bibnamefont
            {Shen}}, \bibinfo {author} {\bibfnamefont {G.~D.}\ \bibnamefont {Gu}},
        \bibinfo {author} {\bibfnamefont {X.~H.}\ \bibnamefont {Chen}}, \ and\
        \bibinfo {author} {\bibfnamefont {Y.}~\bibnamefont {Zhang}},\ }\href
    {\doibase 10.1038/s41586-019-1718-x} {\bibfield  {journal} {\bibinfo
            {journal} {Nature}\ }\textbf {\bibinfo {volume} {575}},\ \bibinfo {pages}
        {156} (\bibinfo {year} {2019})}\BibitemShut {NoStop}%
    \bibitem [{\citenamefont {Can}\ \emph {et~al.}(2021)\citenamefont {Can},
        \citenamefont {Tummuru}, \citenamefont {Day}, \citenamefont {Elfimov},
        \citenamefont {Damascelli},\ and\ \citenamefont {Franz}}]{scbi1}%
    \BibitemOpen
    \bibfield  {author} {\bibinfo {author} {\bibfnamefont {O.}~\bibnamefont
            {Can}}, \bibinfo {author} {\bibfnamefont {T.}~\bibnamefont {Tummuru}},
        \bibinfo {author} {\bibfnamefont {R.~P.}\ \bibnamefont {Day}}, \bibinfo
        {author} {\bibfnamefont {I.}~\bibnamefont {Elfimov}}, \bibinfo {author}
        {\bibfnamefont {A.}~\bibnamefont {Damascelli}}, \ and\ \bibinfo {author}
        {\bibfnamefont {M.}~\bibnamefont {Franz}},\ }\href {\doibase
        10.1038/s41567-020-01142-7} {\bibfield  {journal} {\bibinfo  {journal} {Nat.
                Phys.}\ }\textbf {\bibinfo {volume} {17}},\ \bibinfo {pages} {519} (\bibinfo
        {year} {2021})}\BibitemShut {NoStop}%
    \bibitem [{\citenamefont {Mercado}\ \emph {et~al.}(2022)\citenamefont
        {Mercado}, \citenamefont {Sahoo},\ and\ \citenamefont {Franz}}]{scbi3}%
    \BibitemOpen
    \bibfield  {author} {\bibinfo {author} {\bibfnamefont {A.}~\bibnamefont
            {Mercado}}, \bibinfo {author} {\bibfnamefont {S.}~\bibnamefont {Sahoo}}, \
        and\ \bibinfo {author} {\bibfnamefont {M.}~\bibnamefont {Franz}},\ }\href
    {\doibase 10.1103/PhysRevLett.128.137002} {\bibfield  {journal} {\bibinfo
            {journal} {Phys. Rev. Lett.}\ }\textbf {\bibinfo {volume} {128}},\ \bibinfo
        {pages} {137002} (\bibinfo {year} {2022})}\BibitemShut {NoStop}%
    \bibitem [{\citenamefont {Volkov}\ \emph {et~al.}(2023)\citenamefont {Volkov},
        \citenamefont {Wilson}, \citenamefont {Lucht},\ and\ \citenamefont
        {Pixley}}]{scbi2}%
    \BibitemOpen
    \bibfield  {author} {\bibinfo {author} {\bibfnamefont {P.~A.}\ \bibnamefont
            {Volkov}}, \bibinfo {author} {\bibfnamefont {J.~H.}\ \bibnamefont {Wilson}},
        \bibinfo {author} {\bibfnamefont {K.~P.}\ \bibnamefont {Lucht}}, \ and\
        \bibinfo {author} {\bibfnamefont {J.~H.}\ \bibnamefont {Pixley}},\ }\href
    {\doibase 10.1103/PhysRevB.107.174506} {\bibfield  {journal} {\bibinfo
            {journal} {Phys. Rev. B}\ }\textbf {\bibinfo {volume} {107}},\ \bibinfo
        {pages} {174506} (\bibinfo {year} {2023})}\BibitemShut {NoStop}%
    \bibitem [{\citenamefont {Lu}\ and\ \citenamefont
        {S\'en\'echal}(2022)}]{scbi4}%
    \BibitemOpen
    \bibfield  {author} {\bibinfo {author} {\bibfnamefont {X.}~\bibnamefont
            {Lu}}\ and\ \bibinfo {author} {\bibfnamefont {D.}~\bibnamefont
            {S\'en\'echal}},\ }\href {\doibase 10.1103/PhysRevB.105.245127} {\bibfield
        {journal} {\bibinfo  {journal} {Phys. Rev. B}\ }\textbf {\bibinfo {volume}
            {105}},\ \bibinfo {pages} {245127} (\bibinfo {year} {2022})}\BibitemShut
    {NoStop}%
    \bibitem [{\citenamefont {Song}\ \emph {et~al.}(2022)\citenamefont {Song},
        \citenamefont {Zhang},\ and\ \citenamefont {Vishwanath}}]{scbi5}%
    \BibitemOpen
    \bibfield  {author} {\bibinfo {author} {\bibfnamefont {X.-Y.}\ \bibnamefont
            {Song}}, \bibinfo {author} {\bibfnamefont {Y.-H.}\ \bibnamefont {Zhang}}, \
        and\ \bibinfo {author} {\bibfnamefont {A.}~\bibnamefont {Vishwanath}},\
    }\href {\doibase 10.1103/PhysRevB.105.L201102} {\bibfield  {journal}
        {\bibinfo  {journal} {Phys. Rev. B}\ }\textbf {\bibinfo {volume} {105}},\
        \bibinfo {pages} {L201102} (\bibinfo {year} {2022})}\BibitemShut {NoStop}%
    \bibitem [{\citenamefont {Tummuru}\ \emph {et~al.}(2022)\citenamefont
        {Tummuru}, \citenamefont {Lantagne-Hurtubise},\ and\ \citenamefont
        {Franz}}]{scbi6}%
    \BibitemOpen
    \bibfield  {author} {\bibinfo {author} {\bibfnamefont {T.}~\bibnamefont
            {Tummuru}}, \bibinfo {author} {\bibfnamefont {E.}~\bibnamefont
            {Lantagne-Hurtubise}}, \ and\ \bibinfo {author} {\bibfnamefont
            {M.}~\bibnamefont {Franz}},\ }\href {\doibase 10.1103/PhysRevB.106.014520}
    {\bibfield  {journal} {\bibinfo  {journal} {Phys. Rev. B}\ }\textbf {\bibinfo
            {volume} {106}},\ \bibinfo {pages} {014520} (\bibinfo {year}
        {2022})}\BibitemShut {NoStop}%
    \bibitem [{\citenamefont {Nandkishore}\ \emph {et~al.}(2012)\citenamefont
        {Nandkishore}, \citenamefont {Levitov},\ and\ \citenamefont
        {Chubukov}}]{nandkishore2012chiral}%
    \BibitemOpen
    \bibfield  {author} {\bibinfo {author} {\bibfnamefont {R.}~\bibnamefont
            {Nandkishore}}, \bibinfo {author} {\bibfnamefont {L.~S.}\ \bibnamefont
            {Levitov}}, \ and\ \bibinfo {author} {\bibfnamefont {A.~V.}\ \bibnamefont
            {Chubukov}},\ }\href {https://www.nature.com/articles/nphys2208} {\bibfield
        {journal} {\bibinfo  {journal} {Nat. Phys.}\ }\textbf {\bibinfo {volume}
            {8}},\ \bibinfo {pages} {158} (\bibinfo {year} {2012})}\BibitemShut {NoStop}%
    \bibitem [{\citenamefont {{Black-Schaffer}}(2012)}]{black-schaffer_edge_2012}%
    \BibitemOpen
    \bibfield  {author} {\bibinfo {author} {\bibfnamefont {A.~M.}\ \bibnamefont
            {{Black-Schaffer}}},\ }\href {\doibase 10.1103/PhysRevLett.109.197001}
    {\bibfield  {journal} {\bibinfo  {journal} {Phys. Rev. Lett.}\ }\textbf
        {\bibinfo {volume} {109}},\ \bibinfo {pages} {197001} (\bibinfo {year}
        {2012})}\BibitemShut {NoStop}%
    \bibitem [{\citenamefont {{Black-Schaffer}}\ and\ \citenamefont
        {Le~Hur}(2015)}]{black-schaffer_topological_2015}%
    \BibitemOpen
    \bibfield  {author} {\bibinfo {author} {\bibfnamefont {A.~M.}\ \bibnamefont
            {{Black-Schaffer}}}\ and\ \bibinfo {author} {\bibfnamefont {K.}~\bibnamefont
            {Le~Hur}},\ }\href {\doibase 10.1103/PhysRevB.92.140503} {\bibfield
        {journal} {\bibinfo  {journal} {Phys. Rev. B}\ }\textbf {\bibinfo {volume}
            {92}},\ \bibinfo {pages} {140503} (\bibinfo {year} {2015})}\BibitemShut
    {NoStop}%
    \bibitem [{\citenamefont {Liu}\ \emph {et~al.}(2013)\citenamefont {Liu},
        \citenamefont {Liu}, \citenamefont {Wu}, \citenamefont {Yang},\ and\
        \citenamefont {Yao}}]{liu_d_2013}%
    \BibitemOpen
    \bibfield  {author} {\bibinfo {author} {\bibfnamefont {F.}~\bibnamefont
            {Liu}}, \bibinfo {author} {\bibfnamefont {C.-C.}\ \bibnamefont {Liu}},
        \bibinfo {author} {\bibfnamefont {K.}~\bibnamefont {Wu}}, \bibinfo {author}
        {\bibfnamefont {F.}~\bibnamefont {Yang}}, \ and\ \bibinfo {author}
        {\bibfnamefont {Y.}~\bibnamefont {Yao}},\ }\href {\doibase
        10.1103/PhysRevLett.111.066804} {\bibfield  {journal} {\bibinfo  {journal}
            {Phys. Rev. Lett.}\ }\textbf {\bibinfo {volume} {111}},\ \bibinfo {pages}
        {066804} (\bibinfo {year} {2013})}\BibitemShut {NoStop}%
    \bibitem [{\citenamefont {Yang}\ \emph {et~al.}(2018)\citenamefont {Yang},
        \citenamefont {Qin}, \citenamefont {Zhang}, \citenamefont {Fang},\ and\
        \citenamefont {Hu}}]{yang__2018-1}%
    \BibitemOpen
    \bibfield  {author} {\bibinfo {author} {\bibfnamefont {Z.}~\bibnamefont
            {Yang}}, \bibinfo {author} {\bibfnamefont {S.}~\bibnamefont {Qin}}, \bibinfo
        {author} {\bibfnamefont {Q.}~\bibnamefont {Zhang}}, \bibinfo {author}
        {\bibfnamefont {C.}~\bibnamefont {Fang}}, \ and\ \bibinfo {author}
        {\bibfnamefont {J.}~\bibnamefont {Hu}},\ }\href {\doibase
        10.1103/PhysRevB.98.104515} {\bibfield  {journal} {\bibinfo  {journal} {Phys.
                Rev. B}\ }\textbf {\bibinfo {volume} {98}},\ \bibinfo {pages} {104515}
        (\bibinfo {year} {2018})}\BibitemShut {NoStop}%
    \bibitem [{sup()}]{supp}%
    \BibitemOpen
    \href@noop {} {\bibinfo  {journal} {See Supplemental Material for more
            details about (I) the derivation of the edge Hamiltonian of the Zeeman field
            along the $x$ direction, (II) the edge theory for the {Zeeman} field along
            the $y$ direction, (III) the derivation of the boundary of the phase diagram,
            (IV) effects for the in-plane direction and out-plane direction Zeeman field,
            and (V) the Ginzburg-Landau theory on the chiral $d+id^\prime$ pairing
            superconductivity as the ground state of a twisted bilayer of cuprate
            superconductor with a twist angle of $\theta=\pi/4$, which includes
            {Refs.}~\cite{yandw,wangspm,re65,scbi1,yang__2018-1}}\ }\BibitemShut
    {NoStop}%
    \bibitem [{\citenamefont {Bernevig}\ \emph {et~al.}(2006)\citenamefont
        {Bernevig}, \citenamefont {Hughes},\ and\ \citenamefont
        {Zhang}}]{bernevig_quantum_2006}%
    \BibitemOpen
    \bibfield  {journal} {  }\bibfield  {author} {\bibinfo {author} {\bibfnamefont
            {B.~A.}\ \bibnamefont {Bernevig}}, \bibinfo {author} {\bibfnamefont {T.~L.}\
            \bibnamefont {Hughes}}, \ and\ \bibinfo {author} {\bibfnamefont {S.-C.}\
            \bibnamefont {Zhang}},\ }\href {\doibase 10.1126/science.1133734} {\bibfield
        {journal} {\bibinfo  {journal} {Science}\ }\textbf {\bibinfo {volume}
            {314}},\ \bibinfo {pages} {1757} (\bibinfo {year} {2006})}\BibitemShut
    {NoStop}%
    \bibitem [{\citenamefont {Fu}\ and\ \citenamefont
        {Kane}(2007)}]{fu_topological_2007}%
    \BibitemOpen
    \bibfield  {author} {\bibinfo {author} {\bibfnamefont {L.}~\bibnamefont
            {Fu}}\ and\ \bibinfo {author} {\bibfnamefont {C.~L.}\ \bibnamefont {Kane}},\
    }\href {\doibase 10.1103/PhysRevB.76.045302} {\bibfield  {journal} {\bibinfo
            {journal} {Phys. Rev. B}\ }\textbf {\bibinfo {volume} {76}},\ \bibinfo
        {pages} {045302} (\bibinfo {year} {2007})}\BibitemShut {NoStop}%
    \bibitem [{\citenamefont {Zhu}\ \emph {et~al.}(2021)\citenamefont {Zhu},
        \citenamefont {Liao}, \citenamefont {Zhang}, \citenamefont {Xie},
        \citenamefont {Meng}, \citenamefont {Liu}, \citenamefont {Bai}, \citenamefont
        {Ji}, \citenamefont {Zhang}, \citenamefont {Jiang}, \citenamefont {Zhong},
        \citenamefont {Schneeloch}, \citenamefont {Gu}, \citenamefont {Gu},
        \citenamefont {Ma}, \citenamefont {Zhang},\ and\ \citenamefont
        {Xue}}]{XueTwsited_d-wave2021}%
    \BibitemOpen
    \bibfield  {author} {\bibinfo {author} {\bibfnamefont {Y.}~\bibnamefont
            {Zhu}}, \bibinfo {author} {\bibfnamefont {M.}~\bibnamefont {Liao}}, \bibinfo
        {author} {\bibfnamefont {Q.}~\bibnamefont {Zhang}}, \bibinfo {author}
        {\bibfnamefont {H.-Y.}\ \bibnamefont {Xie}}, \bibinfo {author} {\bibfnamefont
            {F.}~\bibnamefont {Meng}}, \bibinfo {author} {\bibfnamefont {Y.}~\bibnamefont
            {Liu}}, \bibinfo {author} {\bibfnamefont {Z.}~\bibnamefont {Bai}}, \bibinfo
        {author} {\bibfnamefont {S.}~\bibnamefont {Ji}}, \bibinfo {author}
        {\bibfnamefont {J.}~\bibnamefont {Zhang}}, \bibinfo {author} {\bibfnamefont
            {K.}~\bibnamefont {Jiang}}, \bibinfo {author} {\bibfnamefont
            {R.}~\bibnamefont {Zhong}}, \bibinfo {author} {\bibfnamefont
            {J.}~\bibnamefont {Schneeloch}}, \bibinfo {author} {\bibfnamefont
            {G.}~\bibnamefont {Gu}}, \bibinfo {author} {\bibfnamefont {L.}~\bibnamefont
            {Gu}}, \bibinfo {author} {\bibfnamefont {X.}~\bibnamefont {Ma}}, \bibinfo
        {author} {\bibfnamefont {D.}~\bibnamefont {Zhang}}, \ and\ \bibinfo {author}
        {\bibfnamefont {Q.-K.}\ \bibnamefont {Xue}},\ }\href {\doibase
        10.1103/PhysRevX.11.031011} {\bibfield  {journal} {\bibinfo  {journal} {Phys.
                Rev. X}\ }\textbf {\bibinfo {volume} {11}},\ \bibinfo {pages} {031011}
        (\bibinfo {year} {2021})}\BibitemShut {NoStop}%
    \bibitem [{\citenamefont {Zhao}\ \emph {et~al.}(2021)\citenamefont {Zhao},
        \citenamefont {Poccia}, \citenamefont {Cui}, \citenamefont {Volkov},
        \citenamefont {Yoo}, \citenamefont {Engelke}, \citenamefont {Ronen},
        \citenamefont {Zhong}, \citenamefont {Gu}, \citenamefont {Plugge},
        \citenamefont {Tummuru}, \citenamefont {Franz}, \citenamefont {Pixley},\ and\
        \citenamefont {Kim}}]{zhao_emergent_2021}%
    \BibitemOpen
    \bibfield  {author} {\bibinfo {author} {\bibfnamefont {S.~Y.~F.}\
            \bibnamefont {Zhao}}, \bibinfo {author} {\bibfnamefont {N.}~\bibnamefont
            {Poccia}}, \bibinfo {author} {\bibfnamefont {X.}~\bibnamefont {Cui}},
        \bibinfo {author} {\bibfnamefont {P.~A.}\ \bibnamefont {Volkov}}, \bibinfo
        {author} {\bibfnamefont {H.}~\bibnamefont {Yoo}}, \bibinfo {author}
        {\bibfnamefont {R.}~\bibnamefont {Engelke}}, \bibinfo {author} {\bibfnamefont
            {Y.}~\bibnamefont {Ronen}}, \bibinfo {author} {\bibfnamefont
            {R.}~\bibnamefont {Zhong}}, \bibinfo {author} {\bibfnamefont
            {G.}~\bibnamefont {Gu}}, \bibinfo {author} {\bibfnamefont {S.}~\bibnamefont
            {Plugge}}, \bibinfo {author} {\bibfnamefont {T.}~\bibnamefont {Tummuru}},
        \bibinfo {author} {\bibfnamefont {M.}~\bibnamefont {Franz}}, \bibinfo
        {author} {\bibfnamefont {J.~H.}\ \bibnamefont {Pixley}}, \ and\ \bibinfo
        {author} {\bibfnamefont {P.}~\bibnamefont {Kim}},\ }\href
    {http://arxiv.org/abs/2108.13455} {\bibfield  {journal} {\bibinfo  {journal}
            {arXiv:2108.13455}\ } (\bibinfo {year} {2021})}\BibitemShut {NoStop}%
    \bibitem [{\citenamefont {Rosenzweig}\ \emph {et~al.}(2020)\citenamefont
        {Rosenzweig}, \citenamefont {Karakachian}, \citenamefont {Marchenko},
        \citenamefont {K{\"u}ster},\ and\ \citenamefont
        {Starke}}]{rosenzweig_overdoping_2020}%
    \BibitemOpen
    \bibfield  {author} {\bibinfo {author} {\bibfnamefont {P.}~\bibnamefont
            {Rosenzweig}}, \bibinfo {author} {\bibfnamefont {H.}~\bibnamefont
            {Karakachian}}, \bibinfo {author} {\bibfnamefont {D.}~\bibnamefont
            {Marchenko}}, \bibinfo {author} {\bibfnamefont {K.}~\bibnamefont
            {K{\"u}ster}}, \ and\ \bibinfo {author} {\bibfnamefont {U.}~\bibnamefont
            {Starke}},\ }\href {\doibase 10.1103/PhysRevLett.125.176403} {\bibfield
        {journal} {\bibinfo  {journal} {Phys. Rev. Lett.}\ }\textbf {\bibinfo
            {volume} {125}},\ \bibinfo {pages} {176403} (\bibinfo {year}
        {2020})}\BibitemShut {NoStop}%
    \bibitem [{\citenamefont {Wu}\ \emph {et~al.}(2018)\citenamefont {Wu},
        \citenamefont {Fatemi}, \citenamefont {Gibson}, \citenamefont {Watanabe},
        \citenamefont {Taniguchi}, \citenamefont {Cava},\ and\ \citenamefont
        {Jarillo-Herrero}}]{wu_observation_2018}%
    \BibitemOpen
    \bibfield  {author} {\bibinfo {author} {\bibfnamefont {S.}~\bibnamefont
            {Wu}}, \bibinfo {author} {\bibfnamefont {V.}~\bibnamefont {Fatemi}}, \bibinfo
        {author} {\bibfnamefont {Q.~D.}\ \bibnamefont {Gibson}}, \bibinfo {author}
        {\bibfnamefont {K.}~\bibnamefont {Watanabe}}, \bibinfo {author}
        {\bibfnamefont {T.}~\bibnamefont {Taniguchi}}, \bibinfo {author}
        {\bibfnamefont {R.~J.}\ \bibnamefont {Cava}}, \ and\ \bibinfo {author}
        {\bibfnamefont {P.}~\bibnamefont {Jarillo-Herrero}},\ }\href {\doibase
        10.1126/science.aan6003} {\bibfield  {journal} {\bibinfo  {journal}
            {Science}\ }\textbf {\bibinfo {volume} {359}},\ \bibinfo {pages} {76}
        (\bibinfo {year} {2018})}\BibitemShut {NoStop}%
    \bibitem [{\citenamefont {Reis}\ \emph {et~al.}(2017)\citenamefont {Reis},
        \citenamefont {Li}, \citenamefont {Dudy}, \citenamefont {Bauernfeind},
        \citenamefont {Glass}, \citenamefont {Hanke}, \citenamefont {Thomale},
        \citenamefont {Schäfer},\ and\ \citenamefont
        {Claessen}}]{reis_bismuthene_2017}%
    \BibitemOpen
    \bibfield  {author} {\bibinfo {author} {\bibfnamefont {F.}~\bibnamefont
            {Reis}}, \bibinfo {author} {\bibfnamefont {G.}~\bibnamefont {Li}}, \bibinfo
        {author} {\bibfnamefont {L.}~\bibnamefont {Dudy}}, \bibinfo {author}
        {\bibfnamefont {M.}~\bibnamefont {Bauernfeind}}, \bibinfo {author}
        {\bibfnamefont {S.}~\bibnamefont {Glass}}, \bibinfo {author} {\bibfnamefont
            {W.}~\bibnamefont {Hanke}}, \bibinfo {author} {\bibfnamefont
            {R.}~\bibnamefont {Thomale}}, \bibinfo {author} {\bibfnamefont
            {J.}~\bibnamefont {Schäfer}}, \ and\ \bibinfo {author} {\bibfnamefont
            {R.}~\bibnamefont {Claessen}},\ }\href {\doibase 10.1126/science.aai8142}
    {\bibfield  {journal} {\bibinfo  {journal} {Science}\ }\textbf {\bibinfo
            {volume} {357}},\ \bibinfo {pages} {287} (\bibinfo {year}
        {2017})}\BibitemShut {NoStop}%
    \bibitem [{\citenamefont {Lüpke}\ \emph {et~al.}(2020)\citenamefont {Lüpke},
        \citenamefont {Waters}, \citenamefont {de~la Barrera}, \citenamefont {Widom},
        \citenamefont {Mandrus}, \citenamefont {Yan}, \citenamefont {Feenstra},\ and\
        \citenamefont {Hunt}}]{lupke_proximity-induced_2020}%
    \BibitemOpen
    \bibfield  {author} {\bibinfo {author} {\bibfnamefont {F.}~\bibnamefont
            {Lüpke}}, \bibinfo {author} {\bibfnamefont {D.}~\bibnamefont {Waters}},
        \bibinfo {author} {\bibfnamefont {S.~C.}\ \bibnamefont {de~la Barrera}},
        \bibinfo {author} {\bibfnamefont {M.}~\bibnamefont {Widom}}, \bibinfo
        {author} {\bibfnamefont {D.~G.}\ \bibnamefont {Mandrus}}, \bibinfo {author}
        {\bibfnamefont {J.}~\bibnamefont {Yan}}, \bibinfo {author} {\bibfnamefont
            {R.~M.}\ \bibnamefont {Feenstra}}, \ and\ \bibinfo {author} {\bibfnamefont
            {B.~M.}\ \bibnamefont {Hunt}},\ }\href {\doibase 10.1038/s41567-020-0816-x}
    {\bibfield  {journal} {\bibinfo  {journal} {Nat. Phys.}\ }\textbf {\bibinfo
            {volume} {16}},\ \bibinfo {pages} {526} (\bibinfo {year} {2020})}\BibitemShut
    {NoStop}%
    \bibitem [{\citenamefont {Shimamura}\ \emph {et~al.}(2018)\citenamefont
        {Shimamura}, \citenamefont {Sugawara}, \citenamefont {Sucharitakul},
        \citenamefont {Souma}, \citenamefont {Iwaya}, \citenamefont {Nakayama},
        \citenamefont {Trang}, \citenamefont {Yamauchi}, \citenamefont {Oguchi},
        \citenamefont {Kudo}, \citenamefont {Noji}, \citenamefont {Koike},
        \citenamefont {Takahashi}, \citenamefont {Hanaguri},\ and\ \citenamefont
        {Sato}}]{shimamura_ultrathin_2018}%
    \BibitemOpen
    \bibfield  {author} {\bibinfo {author} {\bibfnamefont {N.}~\bibnamefont
            {Shimamura}}, \bibinfo {author} {\bibfnamefont {K.}~\bibnamefont {Sugawara}},
        \bibinfo {author} {\bibfnamefont {S.}~\bibnamefont {Sucharitakul}}, \bibinfo
        {author} {\bibfnamefont {S.}~\bibnamefont {Souma}}, \bibinfo {author}
        {\bibfnamefont {K.}~\bibnamefont {Iwaya}}, \bibinfo {author} {\bibfnamefont
            {K.}~\bibnamefont {Nakayama}}, \bibinfo {author} {\bibfnamefont {C.~X.}\
            \bibnamefont {Trang}}, \bibinfo {author} {\bibfnamefont {K.}~\bibnamefont
            {Yamauchi}}, \bibinfo {author} {\bibfnamefont {T.}~\bibnamefont {Oguchi}},
        \bibinfo {author} {\bibfnamefont {K.}~\bibnamefont {Kudo}}, \bibinfo {author}
        {\bibfnamefont {T.}~\bibnamefont {Noji}}, \bibinfo {author} {\bibfnamefont
            {Y.}~\bibnamefont {Koike}}, \bibinfo {author} {\bibfnamefont
            {T.}~\bibnamefont {Takahashi}}, \bibinfo {author} {\bibfnamefont
            {T.}~\bibnamefont {Hanaguri}}, \ and\ \bibinfo {author} {\bibfnamefont
            {T.}~\bibnamefont {Sato}},\ }\href {\doibase 10.1021/acsnano.8b04869}
    {\bibfield  {journal} {\bibinfo  {journal} {ACS Nano}\ }\textbf {\bibinfo
            {volume} {12}},\ \bibinfo {pages} {10977} (\bibinfo {year}
        {2018})}\BibitemShut {NoStop}%
    \bibitem [{\citenamefont {Wang}\ \emph {et~al.}(2013)\citenamefont {Wang},
        \citenamefont {Ding}, \citenamefont {Fedorov}, \citenamefont {Yao},
        \citenamefont {Li}, \citenamefont {Lv}, \citenamefont {Zhao}, \citenamefont
        {Zhang}, \citenamefont {Xu}, \citenamefont {Schneeloch}, \citenamefont
        {Zhong}, \citenamefont {Ji}, \citenamefont {Wang}, \citenamefont {He},
        \citenamefont {Ma}, \citenamefont {Gu}, \citenamefont {Yao}, \citenamefont
        {Xue}, \citenamefont {Chen},\ and\ \citenamefont {Zhou}}]{wang_fully_2013}%
    \BibitemOpen
    \bibfield  {author} {\bibinfo {author} {\bibfnamefont {E.}~\bibnamefont
            {Wang}}, \bibinfo {author} {\bibfnamefont {H.}~\bibnamefont {Ding}}, \bibinfo
        {author} {\bibfnamefont {A.~V.}\ \bibnamefont {Fedorov}}, \bibinfo {author}
        {\bibfnamefont {W.}~\bibnamefont {Yao}}, \bibinfo {author} {\bibfnamefont
            {Z.}~\bibnamefont {Li}}, \bibinfo {author} {\bibfnamefont {Y.-F.}\
            \bibnamefont {Lv}}, \bibinfo {author} {\bibfnamefont {K.}~\bibnamefont
            {Zhao}}, \bibinfo {author} {\bibfnamefont {L.-G.}\ \bibnamefont {Zhang}},
        \bibinfo {author} {\bibfnamefont {Z.}~\bibnamefont {Xu}}, \bibinfo {author}
        {\bibfnamefont {J.}~\bibnamefont {Schneeloch}}, \bibinfo {author}
        {\bibfnamefont {R.}~\bibnamefont {Zhong}}, \bibinfo {author} {\bibfnamefont
            {S.-H.}\ \bibnamefont {Ji}}, \bibinfo {author} {\bibfnamefont
            {L.}~\bibnamefont {Wang}}, \bibinfo {author} {\bibfnamefont {K.}~\bibnamefont
            {He}}, \bibinfo {author} {\bibfnamefont {X.}~\bibnamefont {Ma}}, \bibinfo
        {author} {\bibfnamefont {G.}~\bibnamefont {Gu}}, \bibinfo {author}
        {\bibfnamefont {H.}~\bibnamefont {Yao}}, \bibinfo {author} {\bibfnamefont
            {Q.-K.}\ \bibnamefont {Xue}}, \bibinfo {author} {\bibfnamefont
            {X.}~\bibnamefont {Chen}}, \ and\ \bibinfo {author} {\bibfnamefont
            {S.}~\bibnamefont {Zhou}},\ }\href {\doibase 10.1038/nphys2744} {\bibfield
        {journal} {\bibinfo  {journal} {Nat. Phys.}\ }\textbf {\bibinfo {volume}
            {9}},\ \bibinfo {pages} {621} (\bibinfo {year} {2013})}\BibitemShut {NoStop}%
    \bibitem [{\citenamefont {Xu}\ \emph {et~al.}(2014)\citenamefont {Xu},
        \citenamefont {Liu}, \citenamefont {Wang}, \citenamefont {Ge}, \citenamefont
        {Liu}, \citenamefont {Yang}, \citenamefont {Chen}, \citenamefont {Liu},
        \citenamefont {Xu}, \citenamefont {Gao}, \citenamefont {Qian}, \citenamefont
        {Zhang},\ and\ \citenamefont {Jia}}]{xu_artificial_2014}%
    \BibitemOpen
    \bibfield  {author} {\bibinfo {author} {\bibfnamefont {J.-P.}\ \bibnamefont
            {Xu}}, \bibinfo {author} {\bibfnamefont {C.}~\bibnamefont {Liu}}, \bibinfo
        {author} {\bibfnamefont {M.-X.}\ \bibnamefont {Wang}}, \bibinfo {author}
        {\bibfnamefont {J.}~\bibnamefont {Ge}}, \bibinfo {author} {\bibfnamefont
            {Z.-L.}\ \bibnamefont {Liu}}, \bibinfo {author} {\bibfnamefont
            {X.}~\bibnamefont {Yang}}, \bibinfo {author} {\bibfnamefont {Y.}~\bibnamefont
            {Chen}}, \bibinfo {author} {\bibfnamefont {Y.}~\bibnamefont {Liu}}, \bibinfo
        {author} {\bibfnamefont {Z.-A.}\ \bibnamefont {Xu}}, \bibinfo {author}
        {\bibfnamefont {C.-L.}\ \bibnamefont {Gao}}, \bibinfo {author} {\bibfnamefont
            {D.}~\bibnamefont {Qian}}, \bibinfo {author} {\bibfnamefont {F.-C.}\
            \bibnamefont {Zhang}}, \ and\ \bibinfo {author} {\bibfnamefont {J.-F.}\
            \bibnamefont {Jia}},\ }\href {\doibase 10.1103/PhysRevLett.112.217001}
    {\bibfield  {journal} {\bibinfo  {journal} {Phys. Rev. Lett.}\ }\textbf
        {\bibinfo {volume} {112}},\ \bibinfo {pages} {217001} (\bibinfo {year}
        {2014})}\BibitemShut {NoStop}%
    \bibitem [{\citenamefont {Zhao}\ \emph {et~al.}(2018)\citenamefont {Zhao},
        \citenamefont {Rachmilowitz}, \citenamefont {Ren}, \citenamefont {Han},
        \citenamefont {Schneeloch}, \citenamefont {Zhong}, \citenamefont {Gu},
        \citenamefont {Wang},\ and\ \citenamefont
        {Zeljkovic}}]{zhao_superconducting_2018}%
    \BibitemOpen
    \bibfield  {author} {\bibinfo {author} {\bibfnamefont {H.}~\bibnamefont
            {Zhao}}, \bibinfo {author} {\bibfnamefont {B.}~\bibnamefont {Rachmilowitz}},
        \bibinfo {author} {\bibfnamefont {Z.}~\bibnamefont {Ren}}, \bibinfo {author}
        {\bibfnamefont {R.}~\bibnamefont {Han}}, \bibinfo {author} {\bibfnamefont
            {J.}~\bibnamefont {Schneeloch}}, \bibinfo {author} {\bibfnamefont
            {R.}~\bibnamefont {Zhong}}, \bibinfo {author} {\bibfnamefont
            {G.}~\bibnamefont {Gu}}, \bibinfo {author} {\bibfnamefont {Z.}~\bibnamefont
            {Wang}}, \ and\ \bibinfo {author} {\bibfnamefont {I.}~\bibnamefont
            {Zeljkovic}},\ }\href {\doibase 10.1103/PhysRevB.97.224504} {\bibfield
        {journal} {\bibinfo  {journal} {Phys. Rev. B}\ }\textbf {\bibinfo {volume}
            {97}},\ \bibinfo {pages} {224504} (\bibinfo {year} {2018})}\BibitemShut
    {NoStop}%
    \bibitem [{\citenamefont {Aivazian}\ \emph {et~al.}(2015)\citenamefont
        {Aivazian}, \citenamefont {Gong}, \citenamefont {Jones}, \citenamefont {Chu},
        \citenamefont {Yan}, \citenamefont {Mandrus}, \citenamefont {Zhang},
        \citenamefont {Cobden}, \citenamefont {Yao},\ and\ \citenamefont
        {Xu}}]{aivazian_magnetic_2015}%
    \BibitemOpen
    \bibfield  {author} {\bibinfo {author} {\bibfnamefont {G.}~\bibnamefont
            {Aivazian}}, \bibinfo {author} {\bibfnamefont {Z.}~\bibnamefont {Gong}},
        \bibinfo {author} {\bibfnamefont {A.~M.}\ \bibnamefont {Jones}}, \bibinfo
        {author} {\bibfnamefont {R.-L.}\ \bibnamefont {Chu}}, \bibinfo {author}
        {\bibfnamefont {J.}~\bibnamefont {Yan}}, \bibinfo {author} {\bibfnamefont
            {D.~G.}\ \bibnamefont {Mandrus}}, \bibinfo {author} {\bibfnamefont
            {C.}~\bibnamefont {Zhang}}, \bibinfo {author} {\bibfnamefont
            {D.}~\bibnamefont {Cobden}}, \bibinfo {author} {\bibfnamefont
            {W.}~\bibnamefont {Yao}}, \ and\ \bibinfo {author} {\bibfnamefont
            {X.}~\bibnamefont {Xu}},\ }\href {\doibase 10.1038/nphys3201} {\bibfield
        {journal} {\bibinfo  {journal} {Nat. Phys.}\ }\textbf {\bibinfo {volume}
            {11}},\ \bibinfo {pages} {148} (\bibinfo {year} {2015})}\BibitemShut
    {NoStop}%
    \bibitem [{\citenamefont {Bi}\ \emph {et~al.}(2018)\citenamefont {Bi},
        \citenamefont {Feng}, \citenamefont {Li}, \citenamefont {Niu}, \citenamefont
        {Wang}, \citenamefont {Shi}, \citenamefont {Yu},\ and\ \citenamefont
        {Wu}}]{bi_spin_2018}%
    \BibitemOpen
    \bibfield  {author} {\bibinfo {author} {\bibfnamefont {R.}~\bibnamefont
            {Bi}}, \bibinfo {author} {\bibfnamefont {Z.}~\bibnamefont {Feng}}, \bibinfo
        {author} {\bibfnamefont {X.}~\bibnamefont {Li}}, \bibinfo {author}
        {\bibfnamefont {J.}~\bibnamefont {Niu}}, \bibinfo {author} {\bibfnamefont
            {J.}~\bibnamefont {Wang}}, \bibinfo {author} {\bibfnamefont {Y.}~\bibnamefont
            {Shi}}, \bibinfo {author} {\bibfnamefont {D.}~\bibnamefont {Yu}}, \ and\
        \bibinfo {author} {\bibfnamefont {X.}~\bibnamefont {Wu}},\ }\href {\doibase
        10.1088/1367-2630/aacbef} {\bibfield  {journal} {\bibinfo  {journal} {New J.
                Phys.}\ }\textbf {\bibinfo {volume} {20}},\ \bibinfo {pages} {063026}
        (\bibinfo {year} {2018})}\BibitemShut {NoStop}%
    \bibitem [{\citenamefont {Jäck}\ \emph {et~al.}(2019)\citenamefont {Jäck},
        \citenamefont {Xie}, \citenamefont {Li}, \citenamefont {Jeon}, \citenamefont
        {Bernevig},\ and\ \citenamefont {Yazdani}}]{re70}%
    \BibitemOpen
    \bibfield  {author} {\bibinfo {author} {\bibfnamefont {B.}~\bibnamefont
            {Jäck}}, \bibinfo {author} {\bibfnamefont {Y.}~\bibnamefont {Xie}}, \bibinfo
        {author} {\bibfnamefont {J.}~\bibnamefont {Li}}, \bibinfo {author}
        {\bibfnamefont {S.}~\bibnamefont {Jeon}}, \bibinfo {author} {\bibfnamefont
            {B.~A.}\ \bibnamefont {Bernevig}}, \ and\ \bibinfo {author} {\bibfnamefont
            {A.}~\bibnamefont {Yazdani}},\ }\href {\doibase 10.1126/science.aax1444}
    {\bibfield  {journal} {\bibinfo  {journal} {Science}\ }\textbf {\bibinfo
            {volume} {364}},\ \bibinfo {pages} {1255} (\bibinfo {year}
        {2019})}\BibitemShut {NoStop}%
    \bibitem [{\citenamefont {Law}\ \emph {et~al.}(2009)\citenamefont {Law},
        \citenamefont {Lee},\ and\ \citenamefont {Ng}}]{law_majorana_2009}%
    \BibitemOpen
    \bibfield  {author} {\bibinfo {author} {\bibfnamefont {K.~T.}\ \bibnamefont
            {Law}}, \bibinfo {author} {\bibfnamefont {P.~A.}\ \bibnamefont {Lee}}, \ and\
        \bibinfo {author} {\bibfnamefont {T.~K.}\ \bibnamefont {Ng}},\ }\href
    {\doibase 10.1103/PhysRevLett.103.237001} {\bibfield  {journal} {\bibinfo
            {journal} {Phys. Rev. Lett.}\ }\textbf {\bibinfo {volume} {103}},\ \bibinfo
        {pages} {237001} (\bibinfo {year} {2009})}\BibitemShut {NoStop}%
    \bibitem [{\citenamefont {Wimmer}\ \emph {et~al.}(2011)\citenamefont {Wimmer},
        \citenamefont {Akhmerov}, \citenamefont {Dahlhaus},\ and\ \citenamefont
        {Beenakker}}]{wimmer_quantum_2011}%
    \BibitemOpen
    \bibfield  {author} {\bibinfo {author} {\bibfnamefont {M.}~\bibnamefont
            {Wimmer}}, \bibinfo {author} {\bibfnamefont {A.~R.}\ \bibnamefont
            {Akhmerov}}, \bibinfo {author} {\bibfnamefont {J.~P.}\ \bibnamefont
            {Dahlhaus}}, \ and\ \bibinfo {author} {\bibfnamefont {C.~W.~J.}\ \bibnamefont
            {Beenakker}},\ }\href {\doibase 10.1088/1367-2630/13/5/053016} {\bibfield
        {journal} {\bibinfo  {journal} {New J. Phys.}\ }\textbf {\bibinfo {volume}
            {13}},\ \bibinfo {pages} {053016} (\bibinfo {year} {2011})}\BibitemShut
    {NoStop}%
    \bibitem [{\citenamefont {Nilsson}\ \emph {et~al.}(2008)\citenamefont
        {Nilsson}, \citenamefont {Akhmerov},\ and\ \citenamefont
        {Beenakker}}]{nilsson_splitting_2008}%
    \BibitemOpen
    \bibfield  {author} {\bibinfo {author} {\bibfnamefont {J.}~\bibnamefont
            {Nilsson}}, \bibinfo {author} {\bibfnamefont {A.~R.}\ \bibnamefont
            {Akhmerov}}, \ and\ \bibinfo {author} {\bibfnamefont {C.~W.~J.}\ \bibnamefont
            {Beenakker}},\ }\href {\doibase 10.1103/PhysRevLett.101.120403} {\bibfield
        {journal} {\bibinfo  {journal} {Phys. Rev. Lett.}\ }\textbf {\bibinfo
            {volume} {101}},\ \bibinfo {pages} {120403} (\bibinfo {year}
        {2008})}\BibitemShut {NoStop}%
    \bibitem [{\citenamefont {Jackiw}\ and\ \citenamefont {Rebbi}(1976)}]{re65}%
    \BibitemOpen
    \bibfield  {author} {\bibinfo {author} {\bibfnamefont {R.}~\bibnamefont
            {Jackiw}}\ and\ \bibinfo {author} {\bibfnamefont {C.}~\bibnamefont {Rebbi}},\
    }\href {\doibase 10.1103/PhysRevD.13.3398} {\bibfield  {journal} {\bibinfo
            {journal} {Phys. Rev. D}\ }\textbf {\bibinfo {volume} {13}},\ \bibinfo
        {pages} {3398} (\bibinfo {year} {1976})}\BibitemShut {NoStop}%
\end{thebibliography}%
\end{document}